\documentclass[10pt,a4paper]{paper}
\usepackage[utf8]{inputenc}
\usepackage{graphicx}
\usepackage{float}
\usepackage{amsmath}
\usepackage{bm}
\usepackage{amssymb}
\usepackage{color}
\usepackage[toc,page]{appendix}
\usepackage{cite}
\usepackage{hyperref}
\usepackage{centernot}
\usepackage{subcaption}
\begin{document}

\title{{\bf  \LARGE Shadows and strong gravitational lensing: a brief review}}

 \author{
{\large Pedro~V.~P. Cunha}$^{1,2}$, {\large Carlos A. R. Herdeiro}$^{1}$
\\
\\
$^{1}$ {\small  Departamento de F\'\i sica da Universidade de Aveiro and CIDMA,} \\
{\small Campus de Santiago, 3810-183 Aveiro, Portugal.}
 \\ \texttt{\small  pintodacunha@tecnico.ulisboa.pt; herdeiro@ua.pt}
 \\
 \\
$^{2}${\small Centro de Astrof\'isica e Gravitaç\~ao - CENTRA, Departamento de F\'isica,} \\ {\small Instituto Superior T\'ecnico - IST, Universidade de Lisboa - UL,}\\ {\small Av. Rovisco Pais 1, 1049-001, Lisboa, Portugal}
}

\maketitle

\begin{abstract}

For ultra compact objects (UCOs), Light Rings (LRs) and Fundamental Photon Orbits (FPOs) play a pivotal role in the theoretical analysis of strong gravitational lensing effects, and of BH shadows in particular. In this short review, specific models are considered to illustrate how FPOs can be useful in order to understand some non-trivial gravitational lensing effects. This paper aims at briefly overviewing the theoretical foundations of these effects, touching also some of the related phenomenology, both in General Relativity (GR) and alternative theories of gravity, hopefully providing some intuition and new insights for the underlying physics, which might be critical when testing the Kerr black hole hypothesis. 

\end{abstract}

\tableofcontents

%\newpage

%%%%%%%%%%%%%%%%%%%%%%%%%%%%%%%%%%%%%%%%%%%%%%%%%%%%%%%%%%%%%%%%%%%%%
\section{Introduction}
%%%%%%%%%%%%%%%%%%%%%%%%%%%%%%%%%%%%%%%%%%%%%%%%%%%%%%%%%%%%%%%%%%%%%

One of the extraordinary predictions of General Relativity (GR) was the bending of light rays due to the spacetime curvature, creating a net effect not too dissimilar from that of a lens~\cite{Eddington:1987tk,Chwolson,Einstein:1956zz,1997Sci...275..184R}. Although the measurement of the bending of light was itself instrumental in establishing GR as a physical theory of the Universe, the prospects of a direct observation of a gravitational lens was considered unlikely at the time of Einstein.\\

The discovery of quasars in the 1960s~\cite{1963Natur.197.1040S} brought major advancements to the field of gravitational lensing. Since these sources are both very distant and bright, they are ideal to observe lensing effects if their line of sight is crossed by a massive object (typically a galaxy). In 1979 the first lensing effect of a distant quasar was recorded~\cite{1979Natur.279..381W}, with similar discoveries being made thereafter~\cite{lensingcat}. However, the largest lensing effects that have been presently observed in astrophysical objects are only of the order of a few tens of arc seconds (see $e.g.$~\cite{2003Natur.426..810I}).\\

By contrast, Ultra-Compact Objects (UCOs) can cause much more extreme local deflections of light. These objects (by definition) possess \textit{light rings} (LRs) and can bend light by an \textit{arbitrarily large angle} (see section~\ref{LRs}). LRs are circular photon orbits, an extreme form of light bending with distinct phenomenological signatures in both the electromagnetic and gravitational waves channels.\\

In the gravitational waves channel, the first events detected by LIGO~\cite{Abbott:2016blz,Abbott:2016nmj,Abbott:2017vtc,Abbott:2017oio,Abbott:2017gyy} actually support the existence of LRs (and hence of UCOs), as the post-merger part of the signal (the ringdown) does not carry the direct signature of an event horizon, but rather that of a LR~\cite{Cardoso:2016rao}. Notice that black holes (BHs) fall within the UCO definition: they are UCOs with a horizon. However, there are other compact objects with a LR that could potentially mimick the observed signal. These \textit{horizonless} UCOs are a far more speculative class of objects, which has been widely discussed in the literature for decades. They are motivated by both classical and quantum conceptual issues related to the existence of an event horizon and of a curvature singularity, whose existence in General Relativity follows from Penrose’s singularity theorem, if matter obeys the null energy condition~\cite{Penrose:1964wq,Penrose:1969pc}. Horizonless UCOs recently attracted renewed interest precisely because of the LIGO detections. However, most of these objects lack a (known) dynamical formation mechanism and, as shown recently in~\cite{Cunha:2017qtt}, physically reasonable horizonless UCOs have potential stability issues. This argument relies on the existence of stable LRs and is briefly discussed in section~\ref{LR-pairs}. In addition, the gravitational lensing of a particular horizonless UCO model is analysed in section~\ref{sec_UCO}.\\

In the electromagnetic channel, LRs and FPOs (which generalize the latter, see section~\ref{LRs}) are also closely connected to an important observable that is being targeted by the Event Horizon Telescope: the BH \textit{shadow}~\cite{Loeb:2013lfa}. Simply, the {shadow} of a BH in a given observation frame is the set of directions in the local sky that would receive light from the event horizon; since the latter is not a source of radiation (at least classically) the shadow actually corresponds to a lack of radiation \cite{Bardeen,Synge}. This concept is hence associated with the BH's light absorption cross-section at high frequencies, if the light rays were traced back in time. In particular, the high frequency limit with no polarization is implicitly assumed throughout most of the paper, with light rays simply following null geodesics. But in section~\ref{sec_plasma} some comments are made to the lensing of light in the presence of plasma.\\

The shadow outline in the sky depends on the gravitational lensing of nearby radiation, thus bearing the fingerprint of the geometry around the BH \cite{Johannsen:2015qca}. This builds a particularly exciting prospect for the use of very large baseline interferometry (VLBI) techniques to resolve the angular scale of the event horizon and corresponding shadow of supermassive BH candidates, namely Sagittarius A* in our galaxy center. A shadow observation would probe the spacetime geometry in the vicinity of the horizon, at least as close as the LR orbits, and consequently would test possible deviations to the expected BH geometry ($i.e.$ the Kerr geometry) in this crucial region~\cite{Loeb:2013lfa}.\\

In special cases for which the geodesic motion is integrable ($e.g.$ Kerr), it is possible to have an analytical closed form for the shadow edge (see section~\ref{Kerr-shadow}). However, generically this is not possible and the null geodesic equations have to be solved numerically. This comprises four second order differential equations, although the existence of spacetime symmetries allows some of the four equations to be simplified to first order. Numerically, instead of evolving the light rays directly from a light source and detect the ones that reach the observer, the most efficient procedure actually requires the propagation of light rays from the observer backward in time and identify their origin \cite{Riazuelo:2015shp}, a method named \textit{backwards ray-tracing}.\\

The paper is organised as follows: in section~\ref{LRs} we establish the theoretical foundations by introducing the concept of a LR and a FPO. These concepts are applied to the discussion of the Kerr shadow in section~\ref{Kerr-shadow}, wherein both the exact curve describing the edge of this shadow and an approximate method to obtain it are presented. Some remarks on the possible influence of a plasma surrounding the BH are also made. The introduction of effective potentials in section~\ref{LRs} also allow discussing the stability of FPOs and defining a topological charge for the LRs; the latter is used to establish a theorem on the number of LRs for UCOs that form smoothly from incomplete gravitational collapse starting from approximately flat spacetime. In the remainder of the paper we discuss non-Kerr shadows and lensing. In section~\ref{hairy-BHs} we discuss non-Kerr shadows in GR focusing on the example of Kerr BHs with boson hair. The role of LRs and FPOs is related to some radical differences that can emerge in this example.  The lensing by horizonless UCOs, illustrated by the case of Proca stars is discussed in section~\ref{sec_UCO}. In this case there is no shadow, but a clear signature of (unstable) LRs remains. A brief consideration of shadows in extensions of GR, focusing on the example of Einstein-dilaton-Gauss-Bonnet, is provided in section~\ref{sec_EdGB}. Final remarks are given in section~\ref{sec-concl}.

%%%%%%%%%%%%%%%%%%%%%%%%%%%%%%%%%%%%%%%%%%%%%%%%%%%%%%%%%
\section{Light rings (LRs) and Fundamental Photon Orbits (FPOs)}\label{LRs}
%%%%%%%%%%%%%%%%%%%%%%%%%%%%%%%%%%%%%%%%%%%%%%%%%%%%%%%%%

LRs are a special class of null geodesics, hereafter defined for spacetimes that possess (at least) two commuting Killing vectors ${\bm \zeta}$, ${\bm\xi}$, with $[{\bm \zeta}, {\bm\xi}]=0$; these are associated respectively to stationarity and axial-symmetry of the spacetime, and are expressed in the symmetry adapted coordinates $t,\varphi$ as ${\bm \zeta}=\partial_t$, ${\bm\xi}=\partial_\varphi$. Any null vector tangent to a LR is spanned by a combination of ${\bm\zeta}$, ${\bm\xi}$, and it thus geometrically anchored to these symmetries. As a curious particular case, \textit{static} LRs are possible in some spacetimes; an example occurs at the onset of formation of an ergotorus~\cite{Grandclement:2016eng}. For a static LR $\bm\zeta$ \textit{alone} is always tangent to the LR orbit.\\

LRs can be classified according to their dynamical stability under perturbations. \textit{Unstable} LRs play an important role in strong gravitational lensing and in the formation of BH shadows. For instance, in the paradigmatic Kerr BH of GR all the LRs are {unstable}. Their existence allows light to encircle the BH any number of times before being either scattered back to infinity or plunged into the BH, embodying a scattering threshold. In particular, from an observation perspective, LRs contribute to the boundary of the Kerr shadow. However, we remark that (in general) LRs are not necessarily connected to a shadow edge, namely if multiple unstable LRs are available, or if horizonless UCOs are considered~\cite{Cunha:2016bjh,Cunha:2017wao}.\\

In contrast to the previous case, \textit{stable} LRs if perturbed can revolve closely to the equilibrium trajectory. Although not as common as their unstable relatives, there are multiple examples in the literature which feature stable LRs, $e.g.$ Boson and Proca stars, Kerr BHs with bosonic hair and even wormholes~\cite{Cunha:2016bjh,Cunha:2017wao,Cardoso:2016rao}. One can anticipate that if the spacetime is perturbed, different modes can accumulate and build-up close to a stable LR position, eventually leading to a backreaction on the spacetime. This intuition was reinforced in a paper by Keir~\cite{Keir:2014oka}, in which the existence of a stable LR sets a decay limit for linear waves, being highly suggestive of a non-linear instability. In fact, as discussed in section~\ref{LR-pairs}, horizonless UCOs that are physically reasonable ($e.g.$ smooth, topologically trivial), must have a \textit{stable} LR and are hence prone to non-linear instabilities~\cite{Cunha:2017qtt}.\\

Despite the close connection between LRs and the shadow edge, the former do not entirely determine the latter. Consider again the Kerr case, wherein geodesic motion is Liouville integrable and separates in Boyer-Lindquist coordinates $(t,r_{BL},\theta, \varphi)$~\cite{Carter:1968rr}. For such coordinates, orbits with a constant $r_{BL}$ exist, known in the literature as \textit{spherical orbits}~\cite{Teo} (see section~\ref{Kerr-shadow}). The subset restricted to the equatorial plane, $i.e.$ the surface of $\mathbb{Z}_2$ reflection invariance, are two LRs with co(counter)-rotation with respect to the BH. These LRs coincide in the Schwarzschild limit at $r_{BL}=3M$, where $M$ is the Arnowitt-Deser-Misner (ADM) mass~\cite{PhysRev.116.1322,Bardeen:1972fi}. The set of spherical orbits completely determines the edge of the Kerr shadow, embodying a scattering threshold similar to LRs.\\

From the viewpoint of an observer which sees the Kerr BH lit by a distant (background) celestial sphere, an increasingly larger number of copies of the whole celestial sphere accumulate as we approach an edge in the observer's sky. This edge, parametrized by observation angles, sets the boundary of the Kerr shadow, with each point of the boundary associated to a particular spherical orbit. We remark that the LRs can only determine two points of the shadow edge, if the observer is on the equatorial plane.\\

As it will become apparent from section~\ref{Kerr-shadow}, a vector tangent to a spherical orbit is not (generically) spanned by ${\bm\zeta}$, ${\bm\xi}$, in contrast to LRs. Hence, despite being the natural generalisations of the latter, spherical orbits are intrinsically a different identity. Moreover, orbital analogues of the spherical orbits can exist for spacetimes other than Kerr, even if the geodesic motion is not integrable (see also~\cite{Grover:2017mhm}). Following previous work~\cite{Cunha:2017eoe,Shipley:2016omi}, these orbital generalisations will be designated as \textit{Fundamental Photon Orbits} (FPOs).\\

Similarly to LRs and Kerr's spherical orbits, FPOs are defined for spacetimes with the Killing vectors ${\bm\zeta}$, ${\bm\xi}$, although they have a more complicated formulation. In particular, notice that Kerr spherical orbits were defined in terms of a ``constant radius'' in Boyer-Lindquist coordinates, which is not an invariant statement.
Moreover, a similar criteria in spacetimes for which separability is unknown is meaningless, since $r_{BL}=const.$ is not preserved by mixing $r_{BL}$ and $\theta$, and no basic property favors a particular coordinate chart. \\

Nevertheless, for generic stationary and axisymmetric spacetimes, one can define FPOs as follows~\cite{Cunha:2017eoe}:\\

\textit{Definition:} let $s(\lambda):\mathbb{R}\to\mathcal{M}$ be an affinely parameterised null geodesic, mapping  the real line to the space-time manifold $\mathcal{M}$. $s(\lambda)$ is a FPO if it is restricted to a compact spatial region -- it is a bound state -- and if there is a value $T>0$ for which $s(\lambda)=s(\lambda+T), \forall\,\lambda\in\mathbb{R}$, 
up to isometries.\\

In short, this definition simply requires that an FPO is periodic on the coordinates $(r,\theta)$, by the coordinate notation of the next section, as $(t,\varphi)$ are connected to Killing vectors.\\

To summarise, FPOs in Kerr are provided by spherical photon orbits, which include LRs as a susbset. All FPOs in Kerr are unstable outside the horizon, but more generically FPOs can also be \textit{stable}, potentially leading to non-trivial spacetime instabilities by analogy with the stable LRs. As discussed in~\cite{Cunha:2017eoe}, FPOs can also be paramount in understanding the detailed structure of more generic BH shadows. For instance, consider section~\ref{hairy-BHs}, wherein the interaction between different unstable FPOs can give rise to non-trivial effects at the level of the shadow edge, namely a cusp.

%%%%%%%%%%%%%%%%%%%%%%%%%%%%%%%%%%%%%%%%%%%%%%%%%%%%%%%%%
\subsection{Effective potentials}\label{eff_poten}
%%%%%%%%%%%%%%%%%%%%%%%%%%%%%%%%%%%%%%%%%%%%%%%%%%%%%%%%%

The LR structure of a given spacetime can be analysed even if the the geodesic motion is not fully integrable. The introduction of effective potentials will be particularly useful for that purpose.\\

Consider a 4-dimensional metric, stationary and axially symmetric, written in quasi-isotropic coordinates $(t,r,\theta,\varphi)$~\cite{Cunha:2016bjh,Cunha:2017qtt}. The coordinates $t,\varphi$ are connected respectively to the commuting azimuthal and stationarity Killing vectors ${\bm\zeta}$, ${\bm\xi}$, with the metric being invariant under the simultaneous reflection $t\to-t$ and $\varphi\to-\varphi$. No reflection symmetry $\mathbb{Z}_2$ is required on the equatorial plane $\theta=\pi/2$, and a gauge condition is chosen in order to have $g_{r\theta}=0$, with both $g_{rr}>0$, $g_{\theta\theta}>0$. In order to prevent closed time-like curves we further require $g_{\varphi\varphi}>0$. Unless otherwise specified, no assumptions are made on the field equations, with the results applying to any metric theory of gravity in which photons follow null geodesics.\\

The Hamiltonian $\mathcal{H}=\frac{1}{2}g^{\mu\nu}p_\mu\,p_\nu=0$ determines the null geodesic flow, where $p_\mu$ denotes the photon's four-momentum. The Killing vectors ${\bm\zeta}$, ${\bm\xi}$ yield the constants of geodesic motion $E\equiv -p_t$ and $L\equiv p_\varphi$, respectively interpreted as the photon's energy and angular momentum at infinity. \\

%%%%%%%

The Hamiltonian can be split into a sum of two parts: a potential term, $V(r,\theta) \leqslant 0$ and a kinetic term, $K\geqslant 0$: $2\mathcal{H}=K+V=0$, where
\[K\equiv g^{rr}{p_r}^2  + g^{\theta\theta}{p_\theta}^2  \]
\begin{equation}V=-\frac{1}{D}\left(E^2g_{\varphi\varphi} + 2E\,L g_{t\varphi} + L^2g_{tt}\right),\label{V-def}\end{equation}
where $D\equiv g^2_{t\varphi}-g_{tt}g_{\varphi\varphi}> 0$.
Since the LR's tangent vector is spanned by ${\bm\zeta}$, ${\bm\xi}$, then at LR $p_r=p_\theta=\dot{p}_\mu=0$, where the dot denotes a derivative with respect to an affine parameter. These equalities can be stated in terms of $V$ alone. In particular, notice that from $\mathcal{H}=0$ we can write:  
\[V=0 \quad \Leftrightarrow \quad K=0 \quad \Leftrightarrow \quad p_r=p_\theta=0.\]
Moreover, Hamilton's equations yield:
\[\dot{p}_\mu=-\left(\partial_\mu g^{rr}p_r^2 + \partial_\mu g^{\theta\theta}p_\theta^2 + \partial_\mu V\right)/2.\]
Combining these relations, one can then conclude that at a LR:
\begin{equation}
\quad V=\nabla V=0 \, . 
\label{lrc}
\end{equation}

The potential $V$ has however the disadvantage of depending on the photon parameters $(E,L)$. Below, an alternative potential is constructed that does not have this issue~\cite{Cunha:2016bjh,Cunha:2017eoe}.\\

One should first realise that $L\neq 0$ at a LR. Indeed, consider by \textit{reductio ad absurdum} that $L=0$ and $E\neq 0$; then by eq.~\eqref{V-def} $V\neq 0$, and the LR requirement is violated by eq.~\eqref{lrc}. We could also consider the case for which both $E=L=0$; however this is also not possible, since the energy of a physical photon must be positive for a local observer, yielding $E>-L\,g_{t\varphi}/g_{\varphi\varphi}$ ~\cite{Cunha:2016bjh}.\\

Since $L\neq 0$ at a LR, it is useful to define the (inverse) impact parameter $\sigma\equiv E/L$. With this parameter, $V$ can be factorized as
$V=-{L^2}g_{\varphi\varphi}(\sigma-H_+)(\sigma-H_-)/D$,
where we have introduced the 2D-potential functions $H_\pm$:
\[H_\pm(r,\theta)\equiv \frac{-g_{t\varphi}\pm\sqrt{D}}{g_{\varphi\varphi}}.\]
In contrast to $V$, these potentials are independent on the parameter $\sigma$, and only depend on the metric elements. Additionally, the condition $V=0$ implies one of the mutually exclusive conditions $\sigma=H_+$ or $\sigma=H_-$ to be true, since $H_\pm - H_\mp=\pm 2{\sqrt{D}}/{g_{\varphi\varphi}}\neq 0$. We remark however, that $\sigma=H_\pm(r,\theta)$ is not actually a constraint on $H_\pm$, but it rather determines the required $\sigma$ in order to have $V=0$, given $(r,\theta)$. \\

The LR conditions (\ref{lrc}) in terms of $H_\pm$ are simply transcribed into the single equation $\nabla H_\pm=0$. In other words, a LR is a \textit{critical point} of the potential $H_\pm$, with the value of the latter only determining the LR impact parameter $\sigma$.\\

The stability of a LR can be inferred by the second derivatives of the potentials. In particular, a LR is stable (unstable) along a coordinate $x^\mu$ if $\partial^2_\mu\,V$ is positive (negative). In terms of $H_\pm$, at a LR this is translated into:
\[\partial_\mu^2V=\pm\left(\frac{2L^2}{\sqrt{D}}\right)\partial_\mu^2H_\pm,\]
$i.e.$ the signs of $\partial_\mu^2V$ and $\pm\partial_\mu^2H_\pm$ coincide. The two eigenvectors of the Hessian matrix of $H_\pm$ determine if the LR is a local extremum (saddle point) if both directions have equal (different) stability. In particular, if both directions are stable, then the LR is stable, whereas the latter is unstable if at least one direction is unstable.

%%%%%%%%%%%%%%%%%%%%%%%%%%%%%%%%%%%%%%%%%%%%%%%%%%%%%%%%%%%%%%%%%%%%%
\subsection{Topological charge of a LR}\label{LR-pairs}
%%%%%%%%%%%%%%%%%%%%%%%%%%%%%%%%%%%%%%%%%%%%%%%%%%%%%%%%%%%%%%%%%%%%%

For a continuous family of spacetimes with the Killing vectors $\bm\zeta$,$\bm\xi$, such as $e.g.$ Boson or Proca stars~\cite{Liebling:2012fv,Kleihaus:2005me,Schunck:1996he}, the number of LRs is not a constant (see~\cite{Cunha:2016bjh,Grandclement:2016eng}). However there is still a LR related topological quantity that is preserved~\cite{Cunha:2017qtt}.\\

Consider the stationary and axially-symmetric spacetimes of section~\ref{eff_poten} and a compact and simply connected region $X$ on the $(r,\theta)$ plane for which the metric is \textit{smooth}. One can define a map $f:(r,\theta)\to \nabla H_\pm$, which maps each point of $X$ with coordinates $(r,\theta)$ to a 2D space $Y_\pm$ parameterised by the components $\partial^iH_\pm$, $i\in\{r,\theta\}$. In particular, a critical point of $\nabla H_\pm$ ($i.e.$ a LR) is mapped to the origin of $Y_\pm$.\\

\begin{figure}[t]
\begin{center}
\includegraphics[width=0.8\textwidth]{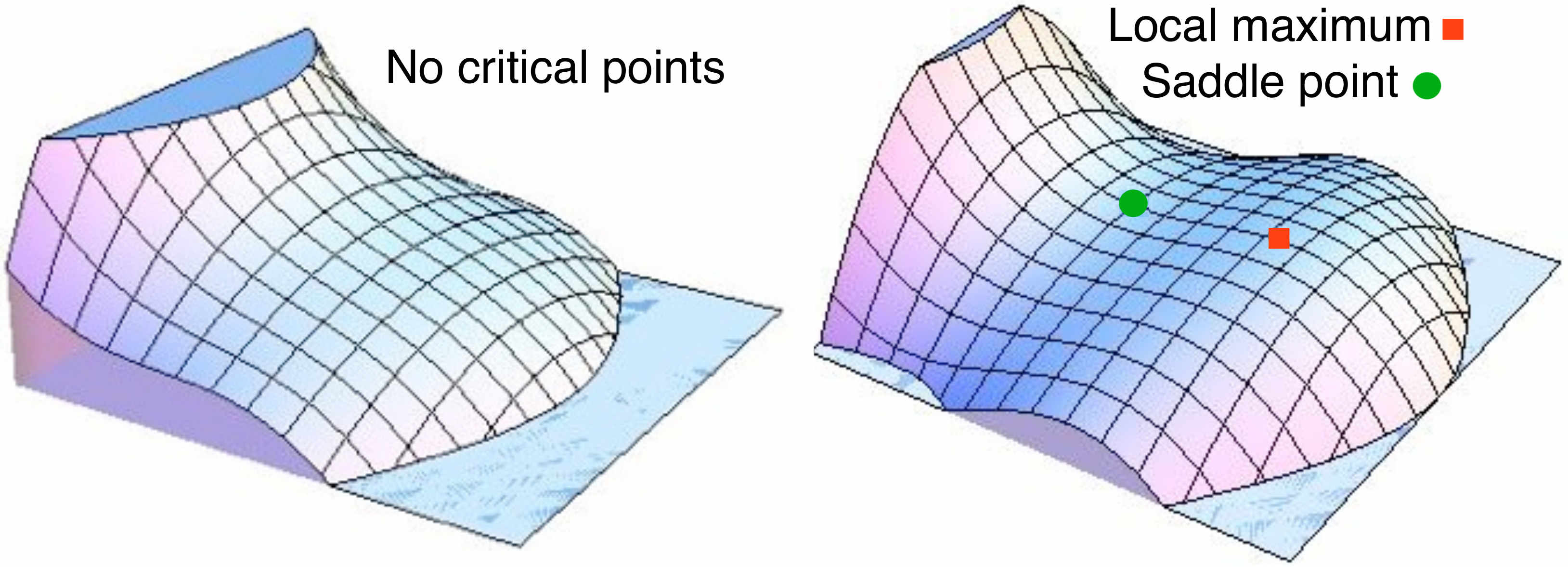}
\caption{\small Conservation of the Brouwer degree under a smooth deformation of a 2D map $(x,y)\rightarrow \nabla H$. We have chosen the illustrative potential $H(x,y)=x(x^2-a)-(1+x^2)y^2$, where $a$ is a \textit{local} deformation parameter that does not affect the asymptotic behavior of the map. Left panel: $a=-2$; there are no critical points and the Brouwer degree is zero. Right panel: $a=1$; there are two critical points, namely one local maximum ($w=+1$) and one saddle point ($w=-1$), with the Brouwer degree still being zero. Adapted from~\cite{Cunha:2017qtt}.}
\label{fig_def}
\end{center}
\end{figure}

Fixing the boundary contribution, one can then compute a topological quantity $w$, called the \textit{Brouwer degree} of the map, that is preserved under smooth deformations of the map ($i.e.$ homotopies)~\cite{steffen2005topology,naber2000topology}. If $\nabla H_\pm=0$ is a regular value of the map, then $w$ can be computed as:
\[w=\sum_k \textrm{sign}(J_k), \qquad J_k=\textrm{det}(\partial_i \partial^j H_\pm)_k,\]
where the sum is over the $k^{th}$ (non-degenerate) LR within the region $X$. In short, one assigns a topological charge $w_k=\pm 1$ to each LR according to the sign of $J_k$, $i.e.$ the Jacobian of the map at the LR location. A \textit{degenerate}\footnote{Unless stated otherwise, the LRs under consideration are non-degenerate. See~\cite{Hod:2017zpi} for a discussion of the degenerate case.} LR corresponds to the limit case in which two LRs with opposite ``charges'' exist at the same location $(r,\theta)$. Due to its topological character, smooth deformations of the metric (and hence of the potentials $H_\pm$) leave the total $w$ preserved. This implies in particular that new LRs are \textit{created in pairs}, with one LR endowed with a $+1$ charge and the other one with a $-1$ (see Fig.~\ref{fig_def} for an illustration).\\

A smooth sequence of solutions within a continuous family of spacetimes can be regarded as a metric deformation, with the assumed symmetries preserved at each stage. However, we remark that, even if a family of solutions is not present, a similar topological argument can still exist.\\

For instance, starting from an approximately flat spacetime, consider an horizonless smooth object that is formed from an incomplete gravitational collapse. Astrophysically, it is reasonable that this final equilibrium state is well described as being stationary, axially-symmetric and asymptotically flat. Moreover, assuming causality, the final state must also be topologically trivial, according to a celebrated theorem by Geroch~\cite{Geroch:1967fs}.\\

In clear contrast to the endpoint states, any intermediate stage of the dynamical collapse is in general neither stationary nor axially-symmetric, unless the collapse process is adiabatic-like. Nevertheless, one can still smoothly deform the endpoint states into each other, via a sequence of \textit{off-shell} spacetimes that possess the Killing vectors $\bm\zeta$,$\bm\xi$. The actual deformation process is irrelevant, being its \textit{existence} that leads to the conclusion that the total $w$ in both the final and initial stages are the same. Since there are no LRs for the initially flat spacetime, $w$ must vanish in both endpoints of the collapse. If our final object has a (non-degenerate) LR ($i.e.$ it is an UCO), then it must possess at least another LR, with a symmetric charge.\\

Furthermore, the stability of each LR can be related to its topological charge. In particular, the analysis of the Jacobian $J=\textrm{det}(\partial_i \partial^j H_\pm)$ leads to the conclusion that a local maximum (or minimum) of $H_\pm$ has $w=+1$, whereas a saddle point of $H_\pm$ has $w=-1$. Similarly, an identical statement in terms of the potential $V$ can also be performed, leading to three types of LRs:
\begin{enumerate}
\item[(a):]{saddle point of $V$ $\longrightarrow$ unstable LR with $w=-1$}
\item[(b):]{local minimum of $V$ $\longrightarrow$ stable LR with $w=+1$}
\item[(c):]{local maximum of $V$ $\longrightarrow$ unstable LR with $w=+1$}
\end{enumerate}

The LRs (a) exist on several spacetimes, namely for the Kerr and Schwarzschild solutions. Moreover, as discussed by~\cite{Cardoso:2016rao}, the ringdown signal of the first LIGO events possess the signature of this LR type.\\

Several spacetimes in the literature also feature LRs of the second category (b), with Proca/Boson stars~\cite{Cunha:2016bjh,Grandclement:2016eng,Cunha:2017wao} or the Majumdar-Papapetrou di-hole system~\cite{Shipley:2016omi} as possible examples. As was previously discussed, these LRs are expected to operate as a radiation trap, leading to a pile up of energy and to an eventual backreaction on the spacetime, possibly triggering a non-linear instability~\cite{Keir:2014oka}.\\

Surprisingly, LRs of the last type (c) are not very frequent. In fact, the authors' are not aware of any literature model featuring this type of LR. Moreover, one can show that the existence of these LRs actually implies a violation of the \textit{Null Energy Condition} (NEC), reason why we shall designate these LRs as \textit{exotic}. The NEC plays a pivotal role in GR, namely being a critical assumption of Penrose's singularity theorem~\cite{Penrose:1964wq,Penrose:1969pc}. Furthermore, the NEC is often considered to be a robust assumption for a healthy theory of gravity, although there can be exceptions~\cite{Rubakov:2014jja}.\\

Assuming Einstein's field equations in geometrized units $G^{\mu\nu}=8\pi T^{\mu\nu}$, the NEC states that $T^{\mu\nu}p_\mu\,p_\nu\geqslant 0$, where $T^{\mu\nu}$ is the energy-momentum tensor and $p_\mu$ is a null vector. Then one can show (see~\cite{Cunha:2017qtt}) that if $p_\mu$ is the LR's four-momentum:
 
\begin{equation}T^{\mu\nu}p_\mu p_\nu =\frac{1}{16\pi}\partial_i\partial^i V,\label{Trace_eq}\end{equation}
which is negative if the LR corresponds to a maximum of V. Hence exotic LRs require a violation of the NEC. However, the converse is not necessarily true, as the NEC can be violated at some point other than the location of the LR. In short:

\[\textrm{Exotic LR}\implies \textrm{NEC violation}\] 
\[\textrm{NEC violation}\centernot\implies  \textrm{Exotic LR}\] 

A similar formulation can hold even in alternative theories of gravity, as long as the field equations can be rewritten as GR with an \textit{effective} energy-momentum tensor, with the NEC now being stated in terms of that tensor. From eq. (\ref{Trace_eq}), one can further conclude that stable LRs are not possible in vacuum, which is consistent with~\cite{Dolan:2016bxj}.\\

In conclusion, if the NEC is enforced, a smooth horizonless UCO that could be a BH mimicker must also possess a stable LR. The latter is then expected to induce a spacetime instability, which possibly creates an issue for these alternative LIGO candidates.\\

As a final remark, let us mention that if similar topological quantities might be defined for generic FPO families, they could be a powerful tool in the analysis of lensing properties.

%%%%%%%%%%%%%%%%%%%%%%%%%%%%%%%%%%%%%%%%%%%%%%%%%%%%%%%%%%%%%%%%%%%%%
\section{The Kerr shadow}\label{Kerr-shadow}
%%%%%%%%%%%%%%%%%%%%%%%%%%%%%%%%%%%%%%%%%%%%%%%%%%%%%%%%%%%%%%%%%%%%%

The Kerr BH is one of the most paradigmatic solutions in GR, having a major potential for astrophysical relevance~\cite{Kerr:1963ud}. Its importance lies on the existence of several uniqueness theorems, which establish that the only stationary, regular, asymptotically flat BH solution of \textit{vacuum} GR is provided by the Kerr metric~\cite{Robinsoon:2004zz,Chrusciel:2012jk,Heusler:1998ua}.\\

Admirably, the null geodesics are fully separable in the Kerr space-time, leading to four constants of motion~\cite{Carter:1968rr}. This allows one to write all four geodesic equations as first order, thus simplifying the problem considerably. Indeed, besides the photon's rest mass (which is zero) and the constants $E,L$ associated to the Killing vectors $\bm\zeta$,$\bm\xi$, one has an additional (hidden) symmetry; the latter is due to existence of a Killing tensor, giving rise to the well-known Carter constant $Q$~\cite{Carter:1968rr}. For null geodesics, the motion dynamics is expressed with just two independent impact parameters:
\[\eta\equiv \frac{L}{E},\qquad \chi=\frac{Q}{E^2}.\]

In Boyer-Lindquist coordinates, Kerr FPOs each exist on a surface with a constant radius $r$, having conveniently been dubbed ``spherical orbits'' in the literature. Notice that while these orbits do lie on surfaces with spherical topology, the geometry is (generically) not spherical. Spherical photon orbits also describe a symmetric motion with respect to the equatorial plane (the surface with $\mathbb{Z}_2$ reflection symmetry) in terms of the Boyer-Lindquist $\theta$ coordinate, reaching a maximum latitude with respect to the symmetry axis.

Given a spherical photon orbit at radius $r$, the corresponding impact parameters must satisfy~\cite{Teo,Wilkins:1972rs}:
\begin{equation}\eta=\frac{r^3+a^2r+Ma^2-3Mr^2}{a(M-r)},\label{eq_eta}\end{equation}
\begin{equation}\chi=\frac{r^2}{r^2-a^2}(3r^2 +a^2 -\eta^2)\geqslant 0,\label{eq_chi}\end{equation}
where $M$ and $a$ are respectively the ADM mass and rotation parameter of the Kerr solution. Moreover, the turning point value $\theta_*$ in a given spherical orbit satisfies:
\begin{equation}a^2\cos^4\theta_*+[\chi+\eta^2-a^2]\cos^2\theta_*-\chi=0.\label{eq_u}\end{equation}

The edge of the Kerr shadow will be the locus of points in the observer's local sky associated to geodesics that barely skim these spherical photon orbits, and hence have the correct values of $\chi,\eta$. The coordinates $(x,y)$ of the Kerr BH shadow edge in the image plane, as seen by a static observer at infinity with coordinate $\theta_o$, are provided by~\cite{Bardeen}:
\begin{equation}x=-\eta/\sin\theta_o,\qquad y=\pm \sqrt{\chi +a^2\cos^2\theta_o -\eta^2/\tan^2\theta_o}\label{eq_shadow}\end{equation}

In this representation, the shadow is defined as a parametric curve, with a dependence on the spherical orbit radius $r$. Hence, for a given value of $r$ we can compute $\{\eta,\chi\}$ and then $\{x,y\}$. The analytical solution for the Kerr shadow appears usually in this way. However, is it possible to write the function $y(x)$ explicitly?

%%%%%%%%%%%%%%%%%%%%%%%%%%%%%%%%%%%%%%%%%%%%%%%%%%%%%%%%%%%%%%%%%%%%%
\subsection{Shadow as a function $y(x)$}
%%%%%%%%%%%%%%%%%%%%%%%%%%%%%%%%%%%%%%%%%%%%%%%%%%%%%%%%%%%%%%%%%%%%%

For an observer at infinity, $\eta$ is trivially obtained from $x$. Also, given $\eta$ and $r$, the value of $\chi$ can be obtained directly from eq. (\ref{eq_chi}). The non-trivial step is only to obtain $r$ given $\eta$. In other words, starting from eq. (\ref{eq_eta}), one has to find the root of this expression:
\[r^3 -3Mr^2 + a(a+\eta) r + Ma(a-\eta)=0.\]
Defining $\mathcal{A}\equiv M^2-\frac{1}{3}a(\eta+a)$ and $\mathcal{B}\equiv M(M^2-a^2)\,\mathcal{|A|}^{-3/2}$, together with Vi{\'e}te's trigonometric trick~\cite{nickalls_2006,nickalls1993new}, one can write the exact (real) solution: 

\begin{align*}
\mathcal{A}>0,\quad \mathcal{B}\leqslant 1: & \qquad\quad r=M+2\sqrt{\mathcal{A}}\cos\left(\frac{1}{3}\arccos\mathcal{B}\right)\\
\mathcal{A}\geqslant 0,\quad \mathcal{B}> 1: & \qquad\quad r=M+2\sqrt{\mathcal{A}}\cosh\left(\frac{1}{3}\log\left[\sqrt{\mathcal{B}^2-1}+\mathcal{B}\right]\right)\\
\mathcal{A}<0: &  \qquad\quad r=M-2\sqrt{|\mathcal{A}|}\sinh\left(\frac{1}{3}\log\left[\sqrt{1+\mathcal{B}^2}-\mathcal{B}\right]\right)
\end{align*}

\begin{figure}[ht]
\begin{center}
\includegraphics[height=6cm]{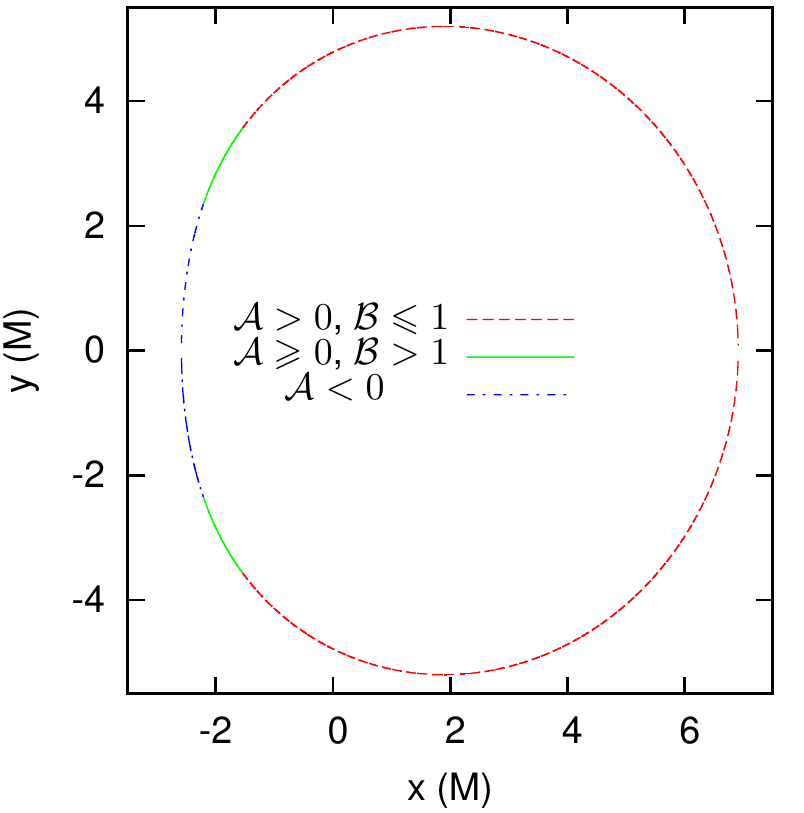}
\end{center}
\caption{\small  Kerr shadow edge function $y(x)$ for $a/M=0.95$. All three branches are necessary to cover the entire edge. The observer is at infinity and in the equatorial plane ($\theta_o=\pi/2$). The axis are in $M$ units.}
\label{fig_shadow-func}
\end{figure}

Hence, given $x$ one can compute $r$ and then $\chi$ and $y$. Notice that each of these branches can describe a different section of the same shadow edge (see Fig. \ref{fig_shadow-func}). This result is consistent with~\cite{Zakharov:2005ek}, since $y(2a)=3\sqrt{3}\,M$ for $\theta_o=\pi/2$.
%

%%%%%%%%%%%%%%%%%%%%%%%%%%%%%%%%%%%%%%%%%%%%%%%%%%%%%%%%%%%%%%%%%%%%%
\subsection{Shadow sketch}
%%%%%%%%%%%%%%%%%%%%%%%%%%%%%%%%%%%%%%%%%%%%%%%%%%%%%%%%%%%%%%%%%%%%%

For the Schwarzschild case, setting $a=0$, we have from equations (\ref{eq_chi})-(\ref{eq_u}):
\[\chi +\eta^2 = 3r^2,\qquad (\chi + \eta^2)\cos^2\theta_*=\chi,\]
whereas from the previous section one concludes that $r=3M$ for all FPOs. For the sake of simplicity, consider a far away observer on the equatorial plane ($\theta_o=\pi/2$), leading to a $y$ shadow coordinate of
\[y=\pm\sqrt{3}\,r\cos\theta_*.\]

Due to spherical symmetry, $r\sin\theta_*=\sqrt{g_{\varphi\varphi}(r,\theta_*)}$, and the expression for $y$ can be re-written in the form:
\begin{equation}y=\pm\sqrt{3}\,\sqrt{g_{\varphi\varphi}(r,\pi/2) - g_{\varphi\varphi}(r,\theta_*)}.\label{eq_sketch}\end{equation}

This is an exact result for Schwarzschild. One can however develop an approximate method to obtain a shadow for other BHs, knowing only the (multiple) radii $r$ at which FPOs occur, their turning points in ``latitude'' and also their impact parameters $\eta$. We critically assume that the contribution of each FPO to the shadow is similar to that of a Schwarzschild spherical orbit in the \textit{same location}.

\begin{figure}[ht]
\begin{center}
\includegraphics[height=6cm]{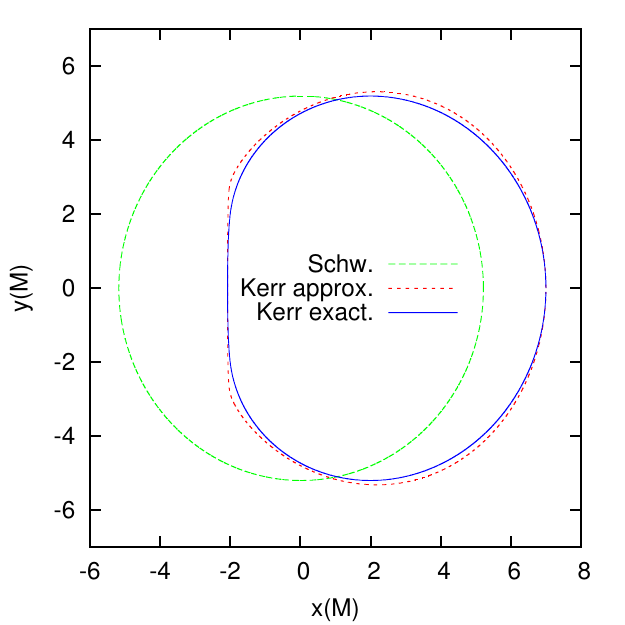}
\end{center}
\caption{\small Shadow of a Schwarzschild BH and a Kerr BH with $a/M=0.999$, together with its approximation. The observer is a at infinity in the equatorial plane. The axis are in $M$ units.}
\label{fig_Kerr}
\end{figure}

Using $x=-\eta$ and equation (\ref{eq_sketch}), we can make a naive prediction for the shadow shape. In particular, we can retry to obtain the Kerr shadow and compare the result with the exact solution (see Fig. \ref{fig_Kerr}). For $a\simeq 0$ the approximation is identical to Schwarzschild, since it is the foundation for the method itself. For the almost extremal case $a\simeq 0.999\,M$ there is not a perfect agreement, but the approximation still manages to capture the main features of the correct shadow, in particular the \textit{D} shape and the horizontal shift. For such a naive calculation, born from spherical symmetry, it is quite surprising. We further remark that this method can be applied with interesting results even for spacetimes that deviate strongly from Kerr, and in which a Carter-like constant is not known, such as Black Holes with scalar hair~\cite{Herdeiro:2015gia,Herdeiro:2014goa}.

%

%%%%%%%%%%%%%%%%%%%%%%%%%%%%%%%%%%%%%%%%%%%%%%%%%%%%%%%%%%%%%%%%%%%%%
\subsection{Light rays in a plasma}
\label{sec_plasma}
%%%%%%%%%%%%%%%%%%%%%%%%%%%%%%%%%%%%%%%%%%%%%%%%%%%%%%%%%%%%%%%%%%%%%

One should not expect astrophysical BHs to exist in total vacuum, but rather surrounded by an accretion disk and ionized matter. Hence, the motion of light rays affected by the presence of a plasma should also be taken into consideration, as the latter could lead to some observable effects. Although this topic has been extensively analysed in the literature ($e.g.$ see~\cite{Muhleman:1970zz,perlick2000ray,Tsupko:2013cqa,Bisnovatyi-Kogan:2015dxa,Abdujabbarov:2016efm,Abdujabbarov:2015pqp}), we shall (very) briefly mention a recent paper by Perlick and Tsupko~\cite{Perlick:2017fio}; the latter contains an interesting discussion on how the Kerr shadow can be modified by the presence of a non-magnetized pressureless plasma, while still keeping the full separability of the geodesic motion. \\

In contrast to vacuum (see section~\ref{eff_poten}), the light propagation in a non-magnetized pressureless plasma is provided by the modified Hamiltonian
\[\mathcal{H}_{p}=\frac{1}{2}\left(g^{\mu\nu}p_\mu p_\nu + \omega_p^2\right)=0,\]
where the plasma frequency $\omega_p$ is proportional to the (square root of the) electron number density for a given point $x^\mu$. Due to the interaction with the plasma, light rays no longer follow null geodesics, as they are endowed with an \textit{effective rest mass}. Indeed, these light rays simply follow \textit{time-like} geodesics under the metric $\widetilde{g}_{\mu\nu}$, which is conformally related to the original one by $\widetilde{g}_{\mu\nu}=\omega^2_p g_{\mu\nu}$~\cite{Perlick:2017fio}.\\

Given a local observer with four-velocity $U^\mu$, normalized to $g_{\mu\nu}U^\mu U^\nu=-1$, the light ray's four-momentum $p^\mu$ can be decomposed into parallel and orthogonal components with respect to $U^\mu$~\cite{Perlick:2017fio}:
\[p^\mu= \omega U^\mu + k^\mu,\]
where the frequency $\omega=-p_\mu U^\mu$ is simply the photon's energy measured in the local observer's frame, and $k^\mu$ is the spacelike wave four-vector. Replacing $p^\mu$ into the Hamiltonian $\mathcal{H}_p$ yields:
\[\omega^2=k_\mu k^\mu + \omega_p^2,\]
which is essentially the relativistic energy balance for a time-like particle. By recalling that the (phase) velocity of the particle is provided by the ratio of its energy by its (linear) momentum, one can compute the index of refraction $n$ of the plasma as:
\[n=\sqrt{1-\frac{\omega^2_p}{\omega^2}}, \qquad \omega^2\geqslant \omega_p^2.\]
Heuristically, the plasma will have an effect similar to a refraction lens, magnifying (or demagnifying) the shadow, depending on the specific model.\\

A generic plasma frequency $\omega_p(x^\mu)$ in the Kerr background will not (in general) preserve the separability of the Hamilton-Jacobi equations, and a Carter-like constant might not exist. As interestingly discussed in~\cite{Perlick:2017fio}, the necessary and sufficient condition on $\omega_p$ for the full integrability of the geodesic equations is, in Boyer-Lindquist coordinates:
\[\omega_p^2(r,\theta)=\frac{f_r(r) + f_\theta(\theta)}{r^2+ a^2\cos^2\theta},\]
where $f_r(r)$ and $f_\theta(\theta)$ are two functions that only depend on $r$ and $\theta$, respectively.

The FPO structure can be significantly modified by the plasma, leading in some cases to a large shadow magnification, to the existence of stable FPOs, and even to a vanishing shadow. However, most of these stronger deviations are only significant if $\omega_p\sim \omega$, which in principle will not be the case in an astrophysical observation.\\

As a curious remark, and despite the breaking of spherical symmetry, the Schwarzschild shadow with a plasma is still a circle regardless of $f_\theta(\theta)$; nevertheless, the shadow size still depends on the observation angle~\cite{Perlick:2017fio}, which is a manifestation of this symmetry breaking.\\

%%%%%%%%%%%%%%%%%%%%%%%%%%%%%%%%%%%%%%%%%%%%%%%%%%%%%%%%%%%%%%%%%%%%%
\section{Non-Kerr shadows in GR}\label{hairy-BHs}
%%%%%%%%%%%%%%%%%%%%%%%%%%%%%%%%%%%%%%%%%%%%%%%%%%%%%%%%%%%%%%%%%%%%%

Due to the uniqueness theorems, the Kerr spacetime is the only physical BH solution in GR, for \textit{vacuum}. However, when considering matter fields, other BH solutions with possible astrophysical relevance can be found. In particular, scalar and Proca fields are some of the simplest matter models one can consider, giving rise to non-trivial BH solutions coupled to these fields. Among these models, \textit{Kerr BHs with bosonic hair} have gathered attention recently, being both physically reasonable and minimally coupled to 4D gravity~\cite{Herdeiro:2014goa,Herdeiro:2015gia,Herdeiro:2016tmi,Cunha:2017eoe}. These BHs are fully non-linear solutions of Einstein's gravity with a complex massive scalar (or Proca) field, moreover being  stationary, axially-symmetric, asymptotically flat and $\mathbb{Z}_2$ symmetric. These solutions exist within GR (and cousin solutions may exist in alternative theories of gravity), they are regular on and outside the horizon, they satisfy all the energy conditions and have no clear pathologies outside the horizon ($e.g.$ close timelike curves or conical singularities). Moreover, Kerr BHs with Proca hair have recently been shown to form dynamically as the endpoint of the superradiant instability, and can thus have a well motivated formation channel~\cite{East:2017ovw,Herdeiro:2017phl}.\\

Kerr BHs with bosonic hair exist within a continuous family of solutions, interpolating between (vacuum) Kerr with a test field~\cite{Hod:2012px,Hod:2013zza} and the corresponding solitonic limit, namely Boson/Proca stars, which do not possess an event horizon. These hairy BHs can possess a surprisingly rich LR and FPO structure, the interplay of which can lead in some cases to unusual effects at the level of the BH shadow and gravitational lensing. We remark that we assume both the scalar and Proca fields to be completely \textit{transparent} to radiation, interacting with light rays only gravitationally.\\

\begin{figure}[ht]
\begin{center}
\includegraphics[height=4.8cm]{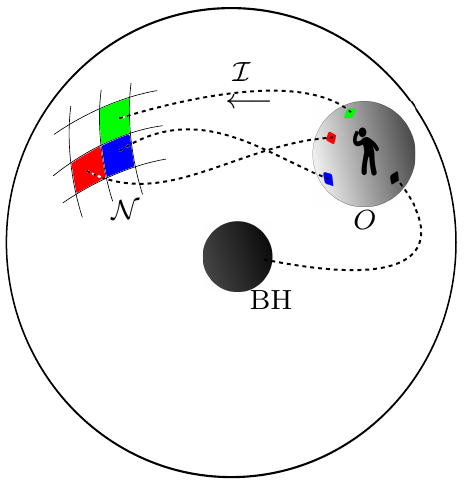}
\end{center}
\caption{\small Schematics of the observational setup. An observer has a local sky $O$, forming a map either to the BH or to the sphere $\mathcal{N}$ surrounding them both.}
\label{fig_setup}
\end{figure}

In the optical channel, the gravitational lensing can strongly modify how an observer perceives its \textit{local sky}. The latter should be interpreted as set of light receiving directions at the location of the observer, being part of the local null tangent space (see also the review~\cite{Perlick:2004tq}). One can make a correspondence between the local sky and a closed $S^2$ manifold ${O}$, parametrized by two observation angles. By placing a light emitting far-away sphere $\mathcal{N}$, surrounding the observer and the BH, some of the light rays will be received in the local sky ${O}$, forming a map $\mathcal{I}:{O}\to \mathcal{N}$, $i.e.$ from $S^2\to S^2$. However, if a BH is present, some points in ${O}$ are actually not mapped to $\mathcal{N}$, as they correspond to light rays that would have originated from the BH. This set of points forms the BH \textit{shadow} (see Fig.~\ref{fig_setup}).\\

In order to represent the map $\mathcal{I}$, and following the setup in~\cite{Bohn:2014xxa,Cunha:2016bjh,Cunha:2015yba}, one first attributes a color to each point in $\mathcal{N}$ according to a regular pattern, say colored quadrants with a grid. Then for each point in $O$ that is not part of the shadow one can compute the color in $\mathcal{N}$ as provided by the map $\mathcal{I}$. The shadow is simply represented in black.\\

\begin{figure}[ht]
\begin{center}
\includegraphics[height=5cm]{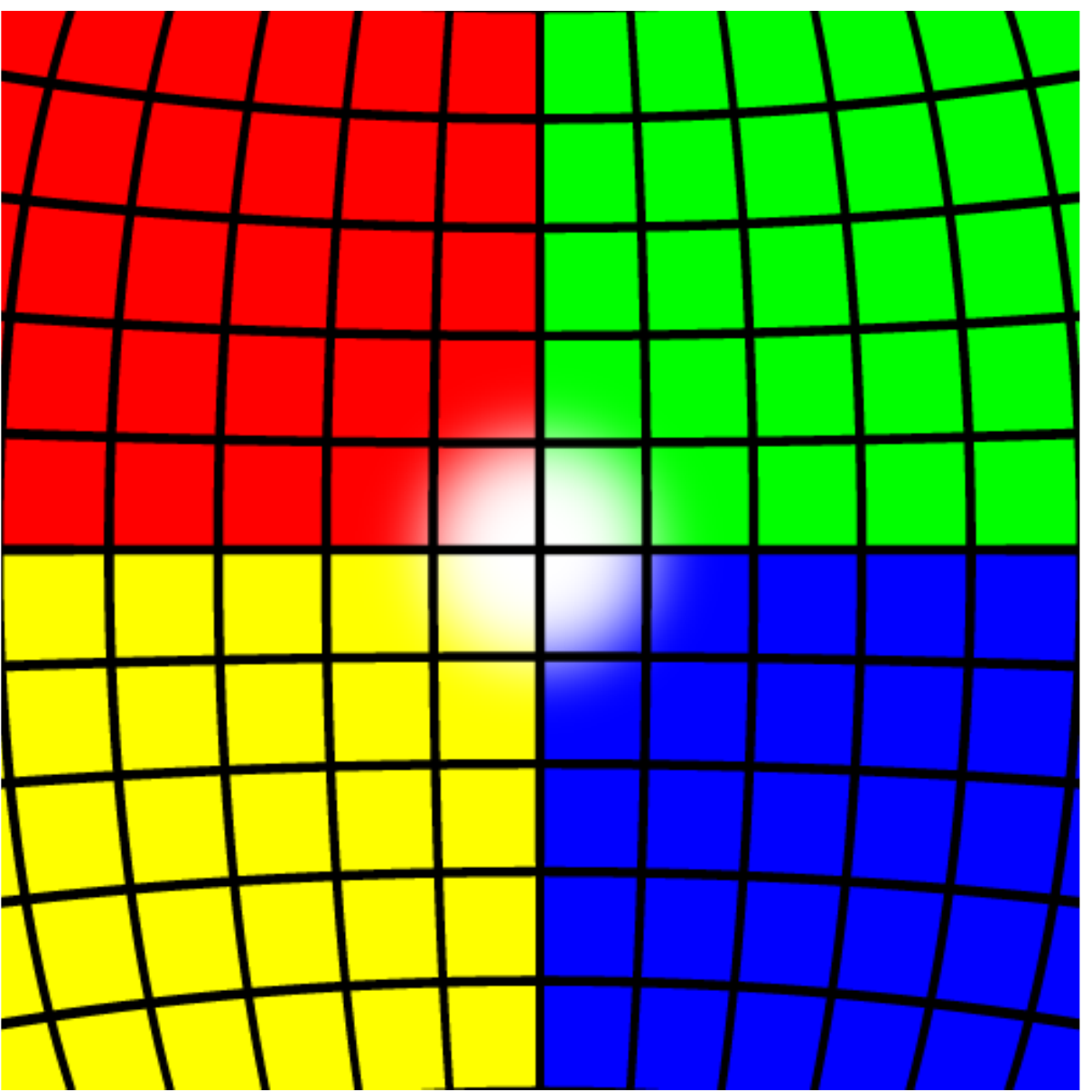}\includegraphics[height=5cm]{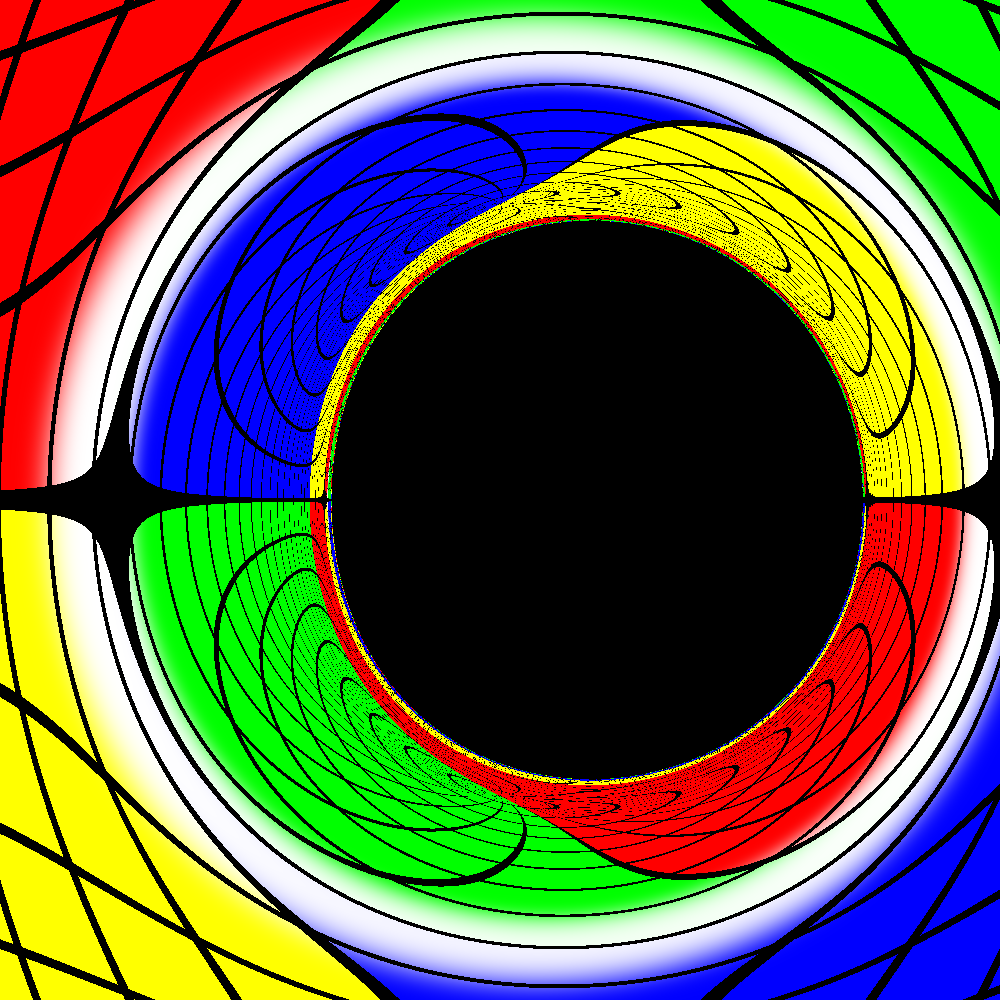}\\
\includegraphics[height=5cm]{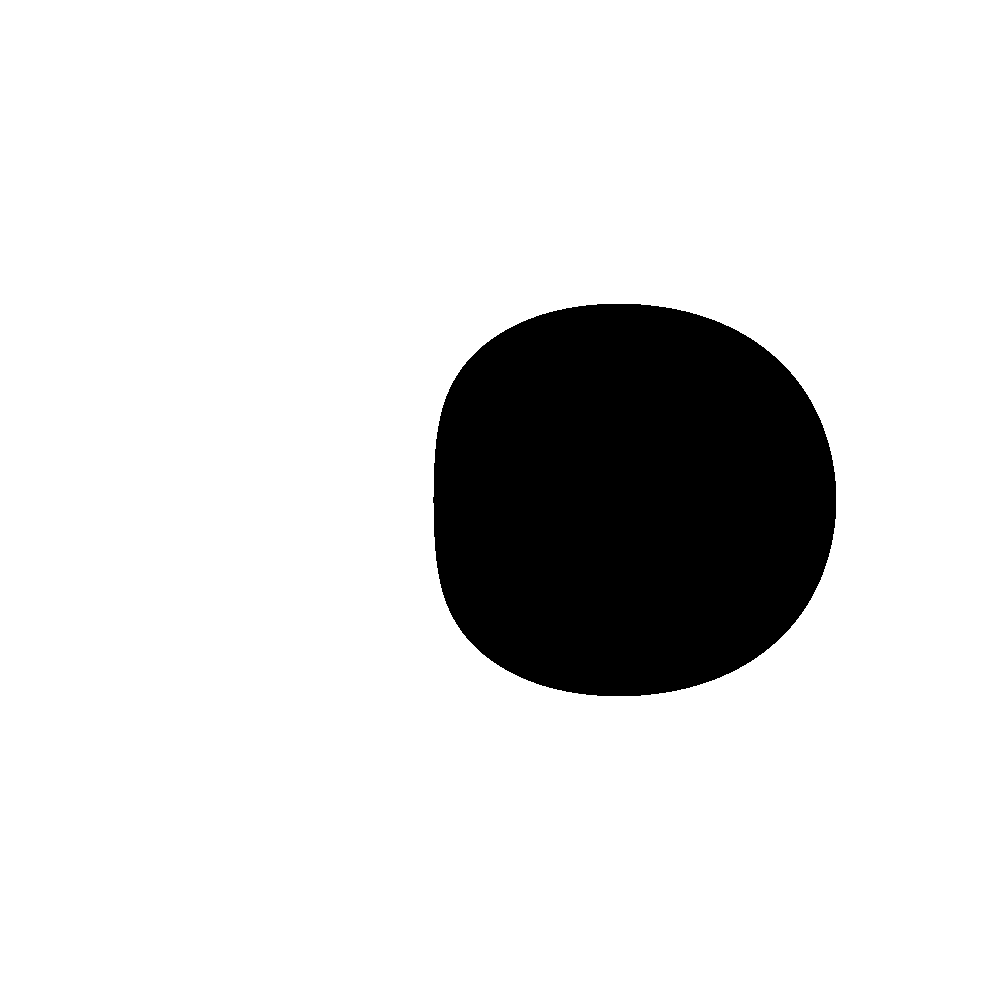}\includegraphics[height=5cm]{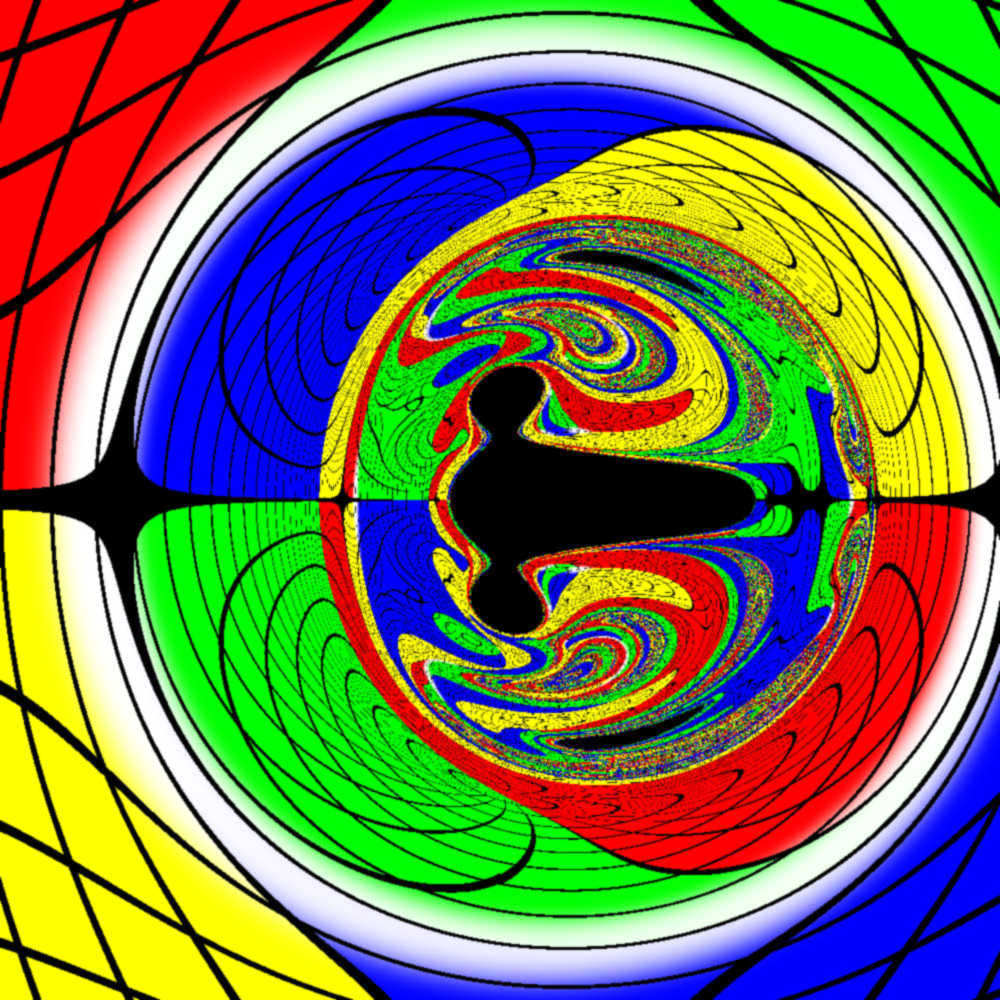}
\end{center}
\caption{\small (From left to right) Observational images in $O$ for \textit{(top row):} flat spacetime; Kerr BH with $a/M\simeq0.82$; \textit{(bottom row):} Kerr-like hairy BH; hairy BH with radical deviations. Adapted from~\cite{Cunha:2015yba}.}
\label{fig_lensing}
\end{figure}

In Fig.~\ref{fig_lensing} comparable sections of $O$ are projected into $\mathbb{R}^2$ \textit{observational images}, not unlike the Cartesian-like plane in Fig.~\ref{fig_Kerr}. In particular, the image's $x$ and $y$-axis represent respectively the azimuthal and latitude coordinates of the local sky $O$, with the origin pointing to the center of the sphere $\mathcal{N}$, where the BH can lie.

The leftmost image of the top row in Fig.~\ref{fig_lensing} displays the observational image in flat spacetime. Since the light rays are not affected by the gravitational field in this case, this image is quite representative of the color pattern in $\mathcal{N}$ that is directly on the line of sight of the observer. In particular, notice that the white dot is in the image center.\\

By placing a Kerr BH in the center of the sphere $\mathcal{N}$ (see top right of Fig.~\ref{fig_lensing}), the white dot is now stretched into a white circle, known as an \textit{Einstein-ring}. Inside the latter one can recognize the Kerr shadow with $a/M\simeq 0.82$, and although it might be unclear from the image, the entire sphere $\mathcal{N}$ is mapped an infinite number of times in-between the Einstein ring and the shadow edge.\\

As previously mentioned, Kerr BHs with bosonic hair have (vacuum) Kerr as one of the endpoints, and so the lensing and shadow might be indistinguishable from the latter. However, if the scalar/Proca field contains a significant fraction\footnote{The mass of the central BH can be determined via Komar integrals.} of the total ADM mass, the observational image can be quite different.\\

Consider for instance the bottom row of Fig.~\ref{fig_lensing}, wherein the leftmost image displays the shadow of a Kerr BH with scalar hair that is still very Kerr-like, with the lensing removed for clarity. In particular, we remark that the shadow has a slightly different shape (it is more squared) and it is also smaller than a comparable\footnote{A comparable Kerr BH has the same ADM mass and angular momentum.} Kerr shadow (see~\cite{Cunha:2016bjh,Cunha:2015yba} for more details). Nevertheless, the FPO structure is still very similar to Kerr.\\

However, as displayed in the rightmost image of the bottom row in Fig.~\ref{fig_lensing}, the shadow of Kerr BHs with scalar hair can be radically different from the Kerr case, both in terms of size, shape and topology~\cite{Cunha:2015yba}. Moreover, the lensing also displays chaotic-like structures, with the latter being connected to the existence of radiation pockets~\cite{Cunha:2016bjh,Shipley:2016omi}. Although not discussed in detail here, the FPO structure of this solution is strikingly different from the Kerr case, which is actually the main reason for these significant differences (for instance, there are four LRs). This hairy BH has almost all of the mass and angular momentum stored in the scalar field, heuristically corresponding to a tiny BH inside a rotating Boson star~\cite{Cunha:2015yba}. Similarly, the FPO structure can also be heuristically regarded as the combination of a Boson Star's FPOs and the FPOs of a central BH. As an illustration of this complex FPO arrangement, notice that there is a circular \textit{ghost shadow edge}, with a Kerr-like profile, surrounding the turbulent part of the image. This is a consequence of a FPO that is not actually responsible for the edge of a shadow, although its lensing signature is still present.\\

%%%%%%%%%%%%%%%%%%%%%%%%%%%%%%%%%%%%%%%%%%%%%%%%%%%%%%%%%
\subsection{Shadow cusp}
%%%%%%%%%%%%%%%%%%%%%%%%%%%%%%%%%%%%%%%%%%%%%%%%%%%%%%%%%

\begin{figure}[ht]
%\begin{subfigure}[b]{.3\linewidth}
%    \centering\includegraphics[height=5.cm]{././contour_PBH.pdf}\vspace*{0.3cm}
% \end{subfigure}%
%\hspace*{1.3cm} 
%\begin{subfigure}[b]{.3\linewidth}
%    \centering\includegraphics[height=5.cm]{././Proca-BH-plot.pdf}
% \end{subfigure}%
%
\begin{center}
\includegraphics[height=5.cm]{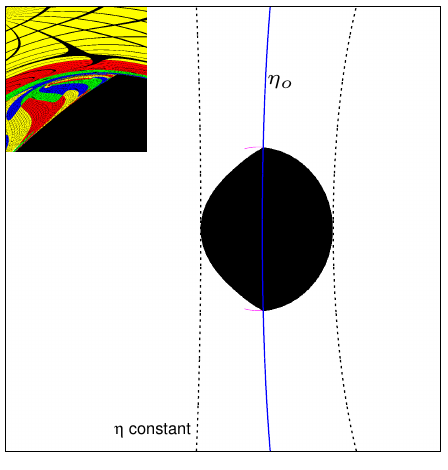}\includegraphics[height=5.cm]{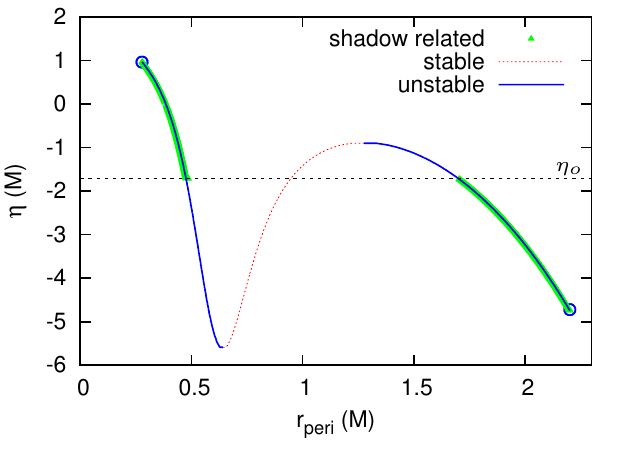}
\end{center}
\caption{\small \textit{Left:} Shadow of a hairy BH with a cusp. The blue line is the set of points with constant $\eta_o$. The inset shows the lensing of a ghost shadow edge (pink curve). \textit{Right:} $\eta$ as a function of the perimetral radius $r_{peri}$ for a continuous FPO family. Notice the branch transition for FPOs related to the shadow edge. Adapted from~\cite{Cunha:2017eoe}.}
\label{fig_cusp}
\end{figure}

In order to illustrate the importance and non-trivial role that FPOs can have at the level of the shadow, consider the leftmost image of Fig.~\ref{fig_cusp}, displaying the shadow of a Kerr BH with Proca hair. In sharp contrast to the previous solutions, the edge of this shadow has a \textit{cusp} and it is thus non-smooth (albeit continuous)~\cite{Cunha:2017eoe}. Surprisingly, this feature can be understood as a consequence of a sharp transition between the FPOs responsible for the shadow edge.\\

As the geodesic motion is not known to be separable, FPOs in this solution generically exist on a surface with non constant $r$ and with non-trivial motion in $\theta$. Additionally, the FPOs relevant for the shadow have a $\mathbb{Z}_2$ reflection symmetry with respect to the equatorial plane $(\theta=\pi/2)$, and each individual FPO intersects this plane at a single radial coordinate $r$.\\

Using this property, one can use the intersection radius as a label for each individual FPO. In particular, the perimetral radius $r_{peri}\equiv \left.\sqrt{g_{\varphi\varphi}}\right|_{_{\theta=\pi/2}}$ , computed at each intersection point, is an invariant quantity related to the $\mathbb{Z}_2$ symmetry and to the Killing vector ${\bm\xi}=\partial_\varphi$. On the right of Fig.~\ref{fig_cusp} the impact parameter $\eta\equiv L/E$ of a continuous FPO family is represented as a function of $r_{peri}$.\\

There are three main branches within this FPO family, two unstable and one stable, with the endpoints being unstable LRs with opposite rotation. A similar FPO diagram also exists for Kerr, although for the latter the intermediate stable branch does not exist, and the FPO $\eta(r_{peri})$ curve has no backbendings.

The thick green line in the right image represents the FPOs that are actually responsible for the shadow edge. There is a sudden transition between the two unstable branches, as marked by the dashed black line for $\eta_o\simeq -1.7M$. This transition coincides with the cusp, as illustrated by the $\eta_o=const.$ blue line in the left of Fig.~\ref{fig_cusp}. Also for the latter, two dotted black lines with constant $\eta$ are represented with the impact parameter of both LRs, each intersecting the shadow at a single point.\\

Still, one can wonder what is the role of the FPOs that are unrelated to the shadow edge. Curiously, these \textit{bare} FPOs that have $\eta< \eta_o$ produce no observable effect, as they are \textit{cloaked} by the shadow being created by FPOs with larger $r_{peri}$. However, (unstable)\footnote{Stable FPOs can also contribute to the lensing despite not producing a sharp signature.} bare FPOs with $\eta>\eta_o$ produce a {ghost} shadow edge, noticeable at the level of the lensing. This is displayed by the pink eyelashes sprouting from the cusp, on the left of Fig.~\ref{fig_cusp}.\\

Similar results have also been reported in~\cite{Wang:2017hjl} for a Konoplya-Zhidenko rotating BH, wherein a transition between spherical orbits leads to a cusp at the level of the shadow. However, in contrast with Kerr BHs with bosonic hair, the geodesic motion is separable in their spacetime.\\

\begin{figure}[ht]
\begin{center}
\includegraphics[height=5cm]{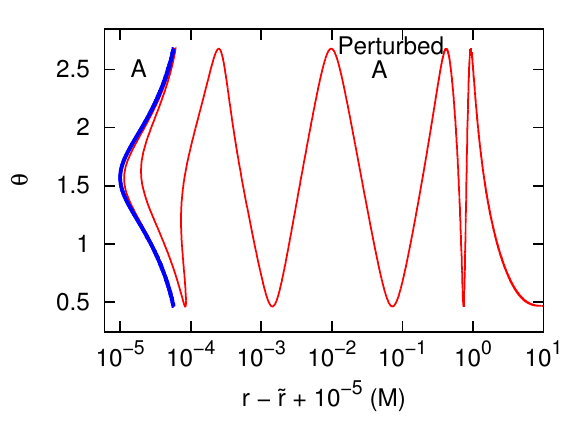}\includegraphics[height=5.cm]{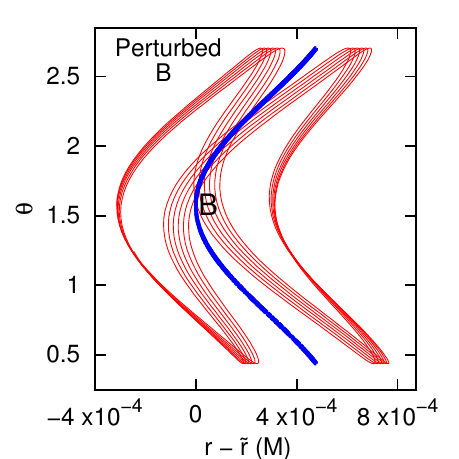}
\end{center}
\caption{\small Projection of two FPOs ($A$ and $B$) on the ($r,\theta$) plane (blue lines). Illustrative perturbations of these orbits are displayed in red, suggesting that $B$ ($A$) is stable (unstable). Adapted from~\cite{Cunha:2017eoe}.}
\label{fig_FPO}
\end{figure}

As a concluding remark, and in order to illustrate the stability properties of FPOs, consider in Fig.~\ref{fig_FPO} two examples of the latter, dubbed $A$ and $B$. These are displayed as blue lines in the figure, together with their perturbed versions in red, to further illustrate their stability. The $x$-axis display the radial coordinates $r$ shifted by $\tilde{r}$, which is respectively the radius at which each FPO intersects the equatorial plane.

The FPO $A$ is represented in the left of Fig.~\ref{fig_FPO}, wherein the $x$-axis has an additional \textit{ad-hoc} radial shift of $10^{-5}$ (notice that the latter is necessary in order to keep all of $A$ visible under the use of a logarithmic scale). The perturbed $A$ orbit is clearly unstable, with the deviation increasing several orders of magnitude in the course of a few oscillations.\\

In contrast to the latter, the FPO $B$ in the right of Fig.~\ref{fig_FPO} appears to be stable, as suggested by its perturbed version. Indeed, the perturbed $B$ orbit never deviates significantly from $B$, simply revolving around the latter as if it was an equilibrium point. We remark that a more precise measure of stability can be made in terms of the Poincaré section of these orbits on the equatorial plane, leading to the same conclusion~\cite{Cunha:2017eoe,jose1998classical}.\\

It is also relevant to mention that the displayed FPOs (in blue) have motion in all coordinates, and in particular these FPOs do not exist at a single $r$ for the chosen coordinate chart. We further stress that a pure FPO is periodic in the ($r,\theta$) plane, $i.e.$ both $A$ and $B$ are always projected to the respective blue lines in the figure, never leaving the latter.

%%%%%%%%%%%%%%%%%%%%%%%%%%%%%%%%%%%%%%%%%%%%%%%%%%%%%%%%%%%%%%%%%%%%%
\section{Lensing by a horizonless UCO}\label{sec_UCO}
%%%%%%%%%%%%%%%%%%%%%%%%%%%%%%%%%%%%%%%%%%%%%%%%%%%%%%%%%%%%%%%%%%%%%

As previously discussed in section~\ref{hairy-BHs}, FPOs can produce sharp effects on the observational image without being connected to the edge of a shadow. This idea will be further reinforced in this section by analysing the gravitational lensing of a particular horizonless UCO: a static Proca star with spherical symmetry~\cite{Brito:2015pxa,Cunha:2017wao}, containing a LR pair with opposite stability.\\

\begin{figure}[ht]
\begin{center}
\includegraphics[height=5.4cm]{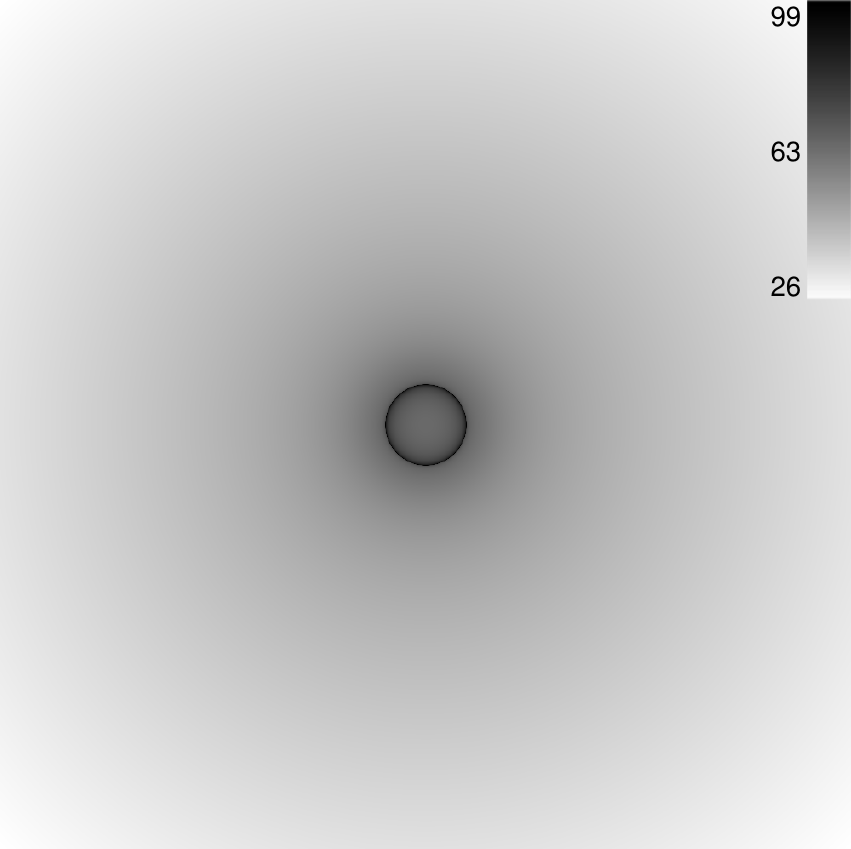}\hspace*{0.2cm}\includegraphics[height=5.4cm]{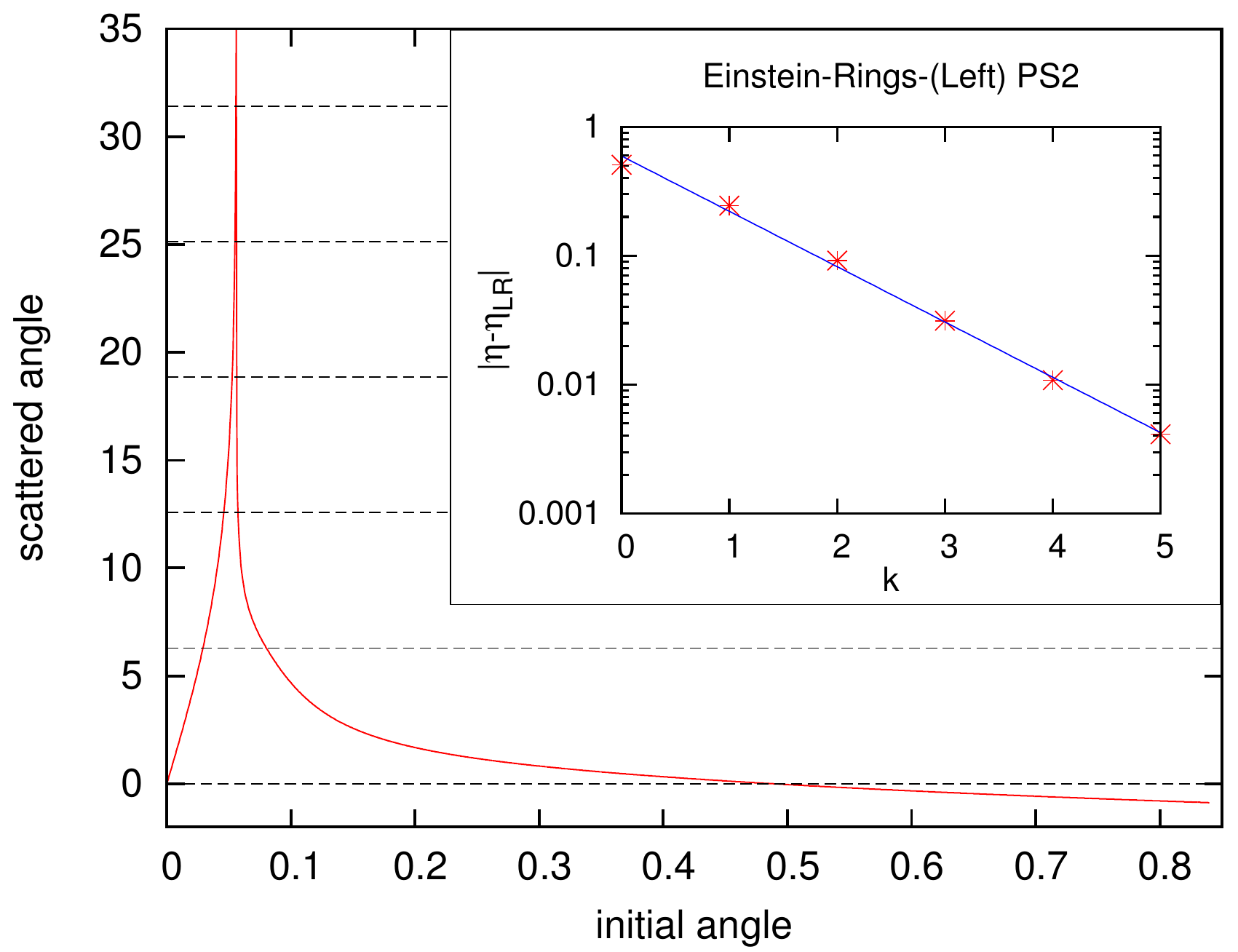}
\end{center}
\caption{\small \textit{Left:} time delay map ($t$ in $M$ units) for a static spherically symmetric Proca star. The darker annular region is a signature of the unstable LR. \textit{Right:} scattered angle as a function of the initial angle; the inset illustrates how well the logarithmic divergence approximates the position of the Einstein rings in the image. Adapted from~\cite{Cunha:2017wao}.}
\label{fig_Proca}
\end{figure}

Consider the left of Fig.~\ref{fig_Proca}, displaying the geodesic \textit{time delay} of the Proca star observational image~\cite{Cunha:2017wao}. This time delay map is similar to the images in Fig.~\ref{fig_lensing}, although the grey levels now represent the variation of the time coordinate $t$ between $\mathcal{N}$ and $O$ (see Fig.~\ref{fig_setup}). This representation sharply reveals an annular region in the sky for which photon motion is much more time consuming. Not too surprisingly, this region is connected to an (unstable)\footnote{The stable LR does not have such a clear lensing signature.} LR orbit.\\

Although there is no event horizon present, and hence no shadow, the attentive reader might notice an uncanny resemblance to a shadow, which is not a coincidence. This particular Proca star has a high density core, leading to a very large redshift of any radiation emitted close to the star's center. In this regard, this configuration is closely related to the concept of a \textit{frozen star}~\cite{1974SvA....17..562P}, the latter being the shadowy afterglow of a star collapsing into a BH, as seen by a faraway observer. Indeed, as discussed in~\cite{Cunha:2017wao}, the fully dynamical evolution of this Proca star quickly leads to a gravitational collapse into a Schwarzschild BH, as this spacetime is plagued with several instabilities (the stable LR might contribute to this). However, despite the resemblance, the angular size of the (final) BH shadow is larger than the (initial) star's annular region, as most of the Proca field mass exists outside the star's unstable LR.\\

Since this Proca star is spherically symmetric, the gravitational lensing can be fully described by a 1D scattering process on the equatorial plane. In particular, the initial angle is provided by the (angular) distance with respect to the observational image center ($i.e.$ in $O$), whereas the scattered angle is the final angle on $\mathcal{N}$, with its origin on the point that would be directly in front of the observer in flat spacetime.\\

The plot on the right of Fig.~\ref{fig_Proca} displays the scattering angle, as a function of the initial angle, with the scattering divergence being a clear signature of the unstable LR. Curiously, the scattering profile for the Schwarzschild BH is quite similar, except for the left part of the peak which would be replaced by the Schwarzschild shadow.\\

Due to symmetry, if the scattered angle is a multiple of $\pi$, then there are points in $O$ along a ring that are mapped to a single point in $\mathcal{N}$, forming a caustic. These rings, commonly known as \textit{Einstein rings}, already appeared in Fig.~\ref{fig_lensing}, with the large white circle being the clearest example. The latter is the lensing of the white point in $\mathcal{N}$ that would be directly in front of the observer in flat spacetime. Hence, any scattering angle multiple of $2\pi$ would lead to such a white circle in $O$. However, a scattering angle of an \textit{odd} multiple of $\pi$ also leads to an Einstein ring, although it corresponds to the lensing of the point in $\mathcal{N}$ that would be directly \textit{behind} the observer. With no loss in generality, we shall focus on the first case. \\

Due to the LR scattering singularity, there is an infinite number of Einstein rings in the image that pile-up close to the LR edge. This LR feature is manifested when representing multiples of $2\pi$ on the right of Fig.~\ref{fig_Proca} using horizontal lines. Moreover, since this divergence of the scattering angle close to the LR is logarithmic, one can write the impact parameter of the $k^{th}$ Einstein ring, corresponding to a scattering angle of $2\pi k$, as:
\[\eta^{(k)}_{ER}\simeq \eta_{LR} + be^{-2\pi k/a},\]
where $\eta_{LR}$ is the impact parameter of the (unstable) LR and $\{a,b\}$ are constants~\cite{Chakraborty:2016lxo}. We remark that, despite not being an angle, the impact parameter $\eta\equiv L/E$ can be used to parametrize the initial angle in $O$ ($e.g.$ see Fig.~\ref{fig_cusp}). In the inset of the right image of Fig.~\ref{fig_Proca}, the numerical values of $|\eta_{ER}-\eta_{LR}|$ are represented as red points, together with the best fit (in blue) to the logarithmic approximation above, showing a good approximation even for the lowest $k$ orders.

%%%%%%%%%%%%%%%%%%%%%%%%%%%%%%%%%%%%%%%%%%%%%%%%%%%%%%%%%%%%%%%%%%%%%
\section{Non-Kerr shadows in alternative theories of gravity}\label{sec_EdGB}
%%%%%%%%%%%%%%%%%%%%%%%%%%%%%%%%%%%%%%%%%%%%%%%%%%%%%%%%%%%%%%%%%%%%%

The discussion in the previous sections only considered spacetimes within GR. However, there are strong theoretical motivations ($e.g.$ non-renormalizability and curvature singularities) to search for alternative theories to Einstein's GR~\cite{Cunha:2016wzk}. Higher order curvature corrections can be included in the action as a simple GR generalization, often leading to field equations with higher order derivatives. Due to covariance, this also leads to time derivatives higher than second order, resulting in unphysical run-away modes (Ostrogradsky instabilities~\cite{Ostrogradsky:1850fid}).\\

Nevertheless, by a cleaver combination of higher curvature terms in the Lagrangian, it is still possible to obtain field equations that are at most second order. In particular, Lovelock~\cite{Lovelock:1971yv} established that in vacuum gravity the most general such combination is provided by the Euler densities $\mathcal{E}_n$, with the latter being scalar polynomial arrangements of the curvature tensor of order $n$. In particular for $D=4$ dimensions, the most general (vacuum) Lovelock theory is a linear combination of $\mathcal{E}_0$ and $\mathcal{E}_1$, simply corresponding to GR with a cosmological constant. Euler densities of higher order, such as the \textit{Gauss-Bonnet} combination $\mathcal{E}_2$, are topological constants in $D=4$, thus not leading to any dynamical contribution when applying the variational method. Nevertheless, by simply coupling $\mathcal{E}_2$ to a dynamical scalar field, the $2^{nd}$ Euler density can generate a non-trivial effect, giving rise to a new theory.\\

The latter model, known in the literature as \textit{Einstein-dilaton-Gauss-Bonnet} (EdGB), occurs naturally as the low energy limit of string theory~\cite{Zwiebach:1985uq} and can also be regarded as an effective description of higher curvature corrections. BHs can be found within the EdGB theory, both in the static~\cite{Kanti:1995vq,Kanti:1997br,Torii:1996yi,Alexeev:1996vs,Melis:2005ji,Chen:2006ge,Chen:2008hk} and rotating case ~\cite{Kleihaus:2011tg,Kleihaus:2015aje,Pani:2009wy,Pani:2011gy, Ayzenberg:2014aka, Maselli:2015tta}. These BH solutions can moreover be perturbatively stable, asymptotically flat and regular, possessing a dilatonic field as a form of non-independent \textit{hair}~\cite{Cunha:2016wzk,Herdeiro:2015waa}. \\

\begin{figure}[ht]
\begin{center}
\includegraphics[height=6.5cm]{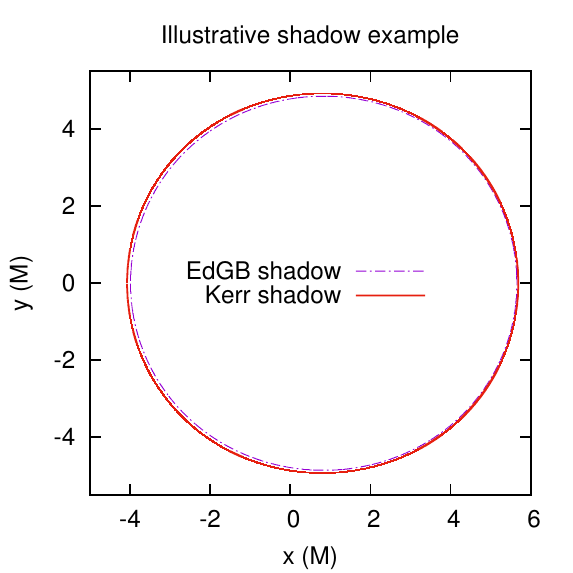}
\end{center}
\caption{\small Shadow of a representative rotating EdGB BH and its comparable Kerr BH ($a/M\simeq 0.41$). Adapted from \cite{Cunha:2016wzk}.}
\label{fig_EdGB}
\end{figure}

We further remark that the Gauss-Bonnet term can be interpreted as an \textit{effective} energy momentum-tensor within plain GR, hence representing some type of \textit{exotic matter} that can violate energy conditions~\cite{Cunha:2016wzk,Herdeiro:2015waa}. One could expect that the distribution of this exotic matter around a EdGB BH would lead to some type of sharp signature at the level of the shadow. However, rather surprisingly, this does not appear to be the case. To illustrate this point, consider Fig.~\ref{fig_EdGB}, wherein the shadow of a rotating EdGB BH is compared with the corresponding Kerr shadow, with the same global ADM quantities. The difference in the shadow size is almost imperceptible (around $\simeq1.4\%$), with the latter being a representative case of most of the EdGB solution space. The main reason for this result appears to be the small variation of the FPO structure with respect to Kerr. Since most of the non-trivial physics exists just outside the horizon, but still enclosed by the FPO structure, any potential new signature from the Gauss-Bonnet term appears to be \textit{hidden} by the BH shadow.\\

This particular model illustrates the fact that new theories of gravity need to significantly modify the LR and FPO structure of the Kerr BH in order to generate a sharp signature at the level of the shadow.

%%%%%%%%%%%%%%%%%%%%%%%%%%%%%%%%%%%%%%%%%%%%%%%%%%%%%%%%%%%%%%%%%%%%%
\section{Conclusions and final remarks}\label{sec-concl}
%%%%%%%%%%%%%%%%%%%%%%%%%%%%%%%%%%%%%%%%%%%%%%%%%%%%%%%%%%%%%%%%%%%%%

Almost 100 years ago, Eddington's observation of light deflection by the sun -- \textit{weak gravitational lensing} -- played a key role in establishing GR as a physical model of the Universe.  With the advent of new observation channels, namely the Event Horizon Telescope, the detection of \textit{strong gravitational lensing} is finally within reach. This prospect has led to a renewed interest, in the XXIst century,  on what is a standard problem in GR: the motion of light around compact objects and in particular the computation of the shadows of BHs. There is already a vast recent literature studying these problems in many different models, see $e.g.$~\cite{Amarilla:2011fx,Yumoto:2012kz,Abdujabbarov:2012bn,Amarilla:2013sj,Nedkova:2013msa,Atamurotov:2013dpa,Atamurotov:2013sca,Li:2013jra,Tinchev:2013nba,Wei:2013kza,Tsukamoto:2014tja,Grenzebach:2014fha,Lu:2014zja,Papnoi:2014aaa,Sakai:2014pga,Psaltis:2014mca,Wei:2015dua,Abdolrahimi:2015rua,Moffat:2015kva,Grenzebach:2015uva,Vincent:2015xta,Grenzebach:2015oea,Abdujabbarov:2015xqa,Ortiz:2015rma,Ghasemi-Nodehi:2015raa,Ohgami:2015nra,Atamurotov:2015xfa, Perlick:2015vta,Bambi:2015rda,Atamurotov:2015nra,Yang:2015hwf,Tinchev:2015apf,Shipley:2016omi,Dolan:2016bxj,Amir:2016cen,Cunha:2015yba,Johannsen:2015hib,Abdujabbarov:2016hnw,Cunha:2016bpi,Huang:2016qnl,Dastan:2016vhb,Younsi:2016azx,Ohgami:2016iqm,Mureika:2016efo,Sharif:2016znp,Tsupko:2017rdo,Bisnovatyi-Kogan:2017kii,Wang:2017hjl, Amir:2017slq,Alhamzawi:2017iyn,Tsukamoto:2017fxq,Mars:2017jkk,Wang:2017qhh,Singh:2017xle,Eiroa:2017uuq}. For \textit{ultra} compact objects (UCOs), Light Rings (LRs) and Fundamental Photon Orbits (FPOs) have a pivotal role in the theoretical analysis of these effects, and of BH shadows in particular. In this brief review, that emphasises these theoretical foundations, specific models were considered  in order to illustrate how FPOs can be instrumental to understand some non-trivial effects at the level of gravitational lensing. This paper aims to be a brief overview and reflection on some of these concepts, substantiated by sharp examples, hopefully providing some intuition and new insights for the underlying physics, which might be critical when testing for the Kerr black hole hypothesis.
  
%%%%%%%%%%%%%%%%%%%%%%%%%%%%%%%%%%%%%%%%%%%%%%%%%%%%%%%%%%%%%%%%%%%%%
\section*{Acknowledgements}
%%%%%%%%%%%%%%%%%%%%%%%%%%%%%%%%%%%%%%%%%%%%%%%%%%%%%%%%%%%%%%%%%%%%%

We would like to thank E. Berti, J. Grover, E. Radu, H. R\'unarsson, A. Wittig for collaboration on some of the work reviewed in this paper. We would also like to thank all the participants in the \textit{Gravitational lensing and black hole shadows workshop} that took place in Aveiro, Portugal, in November 2016, for many stimulating discussions on these topics.  P.C. is supported by Grant No. PD/BD/114071/2015 under the FCT-IDPASC Portugal Ph.D. program. 
C.H. acknowledges funding from the FCT-IF programme. This work was partially supported by the H2020-MSCA-RISE-2015 Grant No. StronGrHEP-690904,
the H2020-MSCA-RISE-2017 Grant No. FunFiCO-777740 and by the CIDMA project UID/MAT/04106/2013 The authors would like to acknowledge networking support by the
COST Action CA16104.
%%%%%%%%%%%%%%%%%%%%%%
%%%   REFERENCES   %%%
%%%%%%%%%%%%%%%%%%%%%%

\bibliography{Ref}{}  

\begin{thebibliography}{100}

\bibitem{Eddington:1987tk}
A.~Eddington, {\em {Space, Time and Gravitation}}.
\newblock Cambridge University Press, 1920.

\bibitem{Chwolson}
O.~Chwolson, ``{\"Uber eine m\"ogliche Form fiktiver Doppelsterne},'' {\em
  Astron. Nachr.}, vol.~221, p.~329, 1924.

\bibitem{Einstein:1956zz}
A.~Einstein, ``{Lens-Like Action of a Star by the Deviation of Light in the
  Gravitational Field},'' {\em Science}, vol.~84, pp.~506--507, 1936.

\bibitem{1997Sci...275..184R}
J.~{Renn}, T.~{Sauer}, and J.~{Stachel}, ``{The origin of gravitational
  lensing: a postscript to Einstein's 1936 Science paper.},'' {\em Science},
  vol.~275, pp.~184--186, Jan. 1997.

\bibitem{1963Natur.197.1040S}
M.~{Schmidt}, ``{3C 273 : A Star-Like Object with Large Red-Shift},'' {\em
  Nature}, vol.~197, p.~1040, Mar. 1963.

\bibitem{1979Natur.279..381W}
D.~{Walsh}, R.~F. {Carswell}, and R.~J. {Weymann}, ``{0957 + 561 A, B - Twin
  quasistellar objects or gravitational lens},'' {\em Nature}, vol.~279,
  pp.~381--384, May 1979.

\bibitem{lensingcat}
``Strong lenses in cosmos.''
\newblock
  http://wwwstaff.ari.uni-heidelberg.de/mitarbeiter/cfaure/cosmos/info.html.

\bibitem{2003Natur.426..810I}
N.~{Inada}, M.~{Oguri}, B.~{Pindor}, J.~F. {Hennawi}, K.~{Chiu}, W.~{Zheng},
  S.-I. {Ichikawa}, M.~D. {Gregg}, R.~H. {Becker}, Y.~{Suto}, M.~A. {Strauss},
  E.~L. {Turner}, C.~R. {Keeton}, J.~{Annis}, F.~J. {Castander}, D.~J.
  {Eisenstein}, J.~A. {Frieman}, M.~{Fukugita}, J.~E. {Gunn}, D.~E. {Johnston},
  S.~M. {Kent}, R.~C. {Nichol}, G.~T. {Richards}, H.-W. {Rix}, E.~S. {Sheldon},
  N.~A. {Bahcall}, J.~{Brinkmann}, {\v Z}.~{Ivezi{\'c}}, D.~Q. {Lamb}, T.~A.
  {McKay}, D.~P. {Schneider}, and D.~G. {York}, ``{A gravitationally lensed
  quasar with quadruple images separated by 14.62arcseconds},'' {\em Nature},
  vol.~426, pp.~810--812, Dec. 2003.

\bibitem{Abbott:2016blz}
B.~P. Abbott {\em et~al.}, ``{Observation of Gravitational Waves from a Binary
  Black Hole Merger},'' {\em Phys. Rev. Lett.}, vol.~116, no.~6, p.~061102,
  2016.

\bibitem{Abbott:2016nmj}
B.~P. Abbott {\em et~al.}, ``{GW151226: Observation of Gravitational Waves from
  a 22-Solar-Mass Binary Black Hole Coalescence},'' {\em Phys. Rev. Lett.},
  vol.~116, no.~24, p.~241103, 2016.

\bibitem{Abbott:2017vtc}
B.~P. Abbott {\em et~al.}, ``{GW170104: Observation of a 50-Solar-Mass Binary
  Black Hole Coalescence at Redshift 0.2},'' {\em Phys. Rev. Lett.}, vol.~118,
  no.~22, p.~221101, 2017.

\bibitem{Abbott:2017oio}
B.~P. Abbott {\em et~al.}, ``{GW170814: A Three-Detector Observation of
  Gravitational Waves from a Binary Black Hole Coalescence},'' {\em Phys. Rev.
  Lett.}, vol.~119, no.~14, p.~141101, 2017.

\bibitem{Abbott:2017gyy}
B.~P. Abbott {\em et~al.}, ``{GW170608: Observation of a 19-solar-mass Binary
  Black Hole Coalescence},'' {\em Astrophys. J.}, vol.~851, no.~2, p.~L35,
  2017.

\bibitem{Cardoso:2016rao}
V.~Cardoso, E.~Franzin, and P.~Pani, ``{Is the gravitational-wave ringdown a
  probe of the event horizon?},'' {\em Phys. Rev. Lett.}, vol.~116, no.~17,
  p.~171101, 2016.
\newblock [Erratum: Phys. Rev. Lett.117,no.8,089902(2016)].

\bibitem{Penrose:1964wq}
R.~Penrose, ``{Gravitational collapse and space-time singularities},'' {\em
  Phys. Rev. Lett.}, vol.~14, pp.~57--59, 1965.

\bibitem{Penrose:1969pc}
R.~Penrose, ``{Gravitational collapse: The role of general relativity},'' {\em
  Riv. Nuovo Cim.}, vol.~1, pp.~252--276, 1969.
\newblock [Gen. Rel. Grav.34,1141(2002)].

\bibitem{Cunha:2017qtt}
P.~V.~P. Cunha, E.~Berti, and C.~A.~R. Herdeiro, ``{Light ring stability in
  ultra-compact objects},'' {\em Phys. Rev. Lett.}, vol.~119, no.~25,
  p.~251102, 2017.

\bibitem{Loeb:2013lfa}
A.~E. Broderick, T.~Johannsen, A.~Loeb, and D.~Psaltis, ``{Testing the No-Hair
  Theorem with Event Horizon Telescope Observations of Sagittarius A*},'' {\em
  Astrophys.J.}, vol.~784, p.~7, 2014.

\bibitem{Bardeen}
J.~M. {Bardeen}, ``{Timelike and null geodesics in the Kerr metric.},'' in {\em
  Black Holes (Les Astres Occlus)} (C.~{Dewitt} and B.~S. {Dewitt}, eds.),
  pp.~215--239, 1973.

\bibitem{Synge}
J.~L. Synge, ``The escape of photons from gravitationally intense stars,'' {\em
  Monthly Notices of the Royal Astronomical Society}, vol.~131, no.~3,
  pp.~463--466, 1966.

\bibitem{Johannsen:2015qca}
T.~Johannsen, ``{Photon Rings around Kerr and Kerr-like Black Holes},'' {\em
  Astrophys.J.}, vol.~777, p.~170, 2013.

\bibitem{Riazuelo:2015shp}
A.~Riazuelo, ``{Seeing relativity -- I. Basics of a raytracing code in a
  Schwarzschild metric},'' 2015.

\bibitem{Grandclement:2016eng}
P.~Grandcl{\'e}ment, ``{Light rings and light points of boson stars},'' {\em
  Phys. Rev.}, vol.~D95, no.~8, p.~084011, 2017.

\bibitem{Cunha:2016bjh}
P.~V.~P. Cunha, J.~Grover, C.~Herdeiro, E.~Radu, H.~Runarsson, and A.~Wittig,
  ``{Chaotic lensing around boson stars and Kerr black holes with scalar
  hair},'' {\em Phys. Rev.}, vol.~D94, no.~10, p.~104023, 2016.

\bibitem{Cunha:2017wao}
P.~V.~P. Cunha, J.~A. Font, C.~Herdeiro, E.~Radu, N.~Sanchis-Gual, and
  M.~Zilhão, ``{Lensing and dynamics of ultracompact bosonic stars},'' {\em
  Phys. Rev.}, vol.~D96, no.~10, p.~104040, 2017.

\bibitem{Keir:2014oka}
J.~Keir, ``{Slowly decaying waves on spherically symmetric spacetimes and
  ultracompact neutron stars},'' {\em Class. Quant. Grav.}, vol.~33, no.~13,
  p.~135009, 2016.

\bibitem{Carter:1968rr}
B.~Carter, ``{Global structure of the Kerr family of gravitational fields},''
  {\em Phys. Rev.}, vol.~174, pp.~1559--1571, 1968.

\bibitem{Teo}
E.~Teo, ``Spherical photon orbits around a kerr black hole,'' {\em General
  Relativity and Gravitation}, vol.~35, no.~11, pp.~1909--1926, 2003.

\bibitem{PhysRev.116.1322}
R.~Arnowitt, S.~Deser, and C.~W. Misner, ``Dynamical structure and definition
  of energy in general relativity,'' {\em Phys. Rev.}, vol.~116,
  pp.~1322--1330, Dec 1959.

\bibitem{Bardeen:1972fi}
J.~M. Bardeen, W.~H. Press, and S.~A. Teukolsky, ``{Rotating black holes:
  Locally nonrotating frames, energy extraction, and scalar synchrotron
  radiation},'' {\em Astrophys.J.}, vol.~178, p.~347, 1972.

\bibitem{Grover:2017mhm}
J.~Grover and A.~Wittig, ``{Black Hole Shadows and Invariant Phase Space
  Structures},'' {\em Phys. Rev.}, vol.~D96, no.~2, p.~024045, 2017.

\bibitem{Cunha:2017eoe}
P.~V.~P. Cunha, C.~A.~R. Herdeiro, and E.~Radu, ``{Fundamental photon orbits:
  black hole shadows and spacetime instabilities},'' {\em Phys. Rev.},
  vol.~D96, no.~2, p.~024039, 2017.

\bibitem{Shipley:2016omi}
J.~Shipley and S.~R. Dolan, ``{Binary black hole shadows, chaotic scattering
  and the Cantor set},'' {\em Class. Quant. Grav.}, vol.~33, no.~17, p.~175001,
  2016.

\bibitem{Liebling:2012fv}
S.~L. Liebling and C.~Palenzuela, ``{Dynamical Boson Stars},'' {\em Living
  Rev.Rel.}, vol.~15, p.~6, 2012.

\bibitem{Kleihaus:2005me}
B.~Kleihaus, J.~Kunz, and M.~List, ``{Rotating boson stars and Q-balls},'' {\em
  Phys.Rev.}, vol.~D72, p.~064002, 2005.

\bibitem{Schunck:1996he}
F.~E. Schunck and E.~W. Mielke, ``{Rotating boson star as an effective mass
  torus in general relativity},'' {\em Phys.Lett.}, vol.~A249, pp.~389--394,
  1998.

\bibitem{steffen2005topology}
 {\em Topology and Geometry in Physics}, pp.~edited by E. Bick and F. D.
  Steffen (Springer, Berlin, Heidelberg, 2005).

\bibitem{naber2000topology}
G.~L. Naber, {\em Topology, geometry, and gauge fields}.
\newblock Springer, New York, 2000.

\bibitem{Hod:2017zpi}
S.~Hod, ``{On the number of light rings in curved spacetimes of ultra-compact
  objects},'' {\em Phys. Lett.}, vol.~B776, pp.~1--4, 2018.

\bibitem{Geroch:1967fs}
R.~P. Geroch, ``{Topology in general relativity},'' {\em J. Math. Phys.},
  vol.~8, pp.~782--786, 1967.

\bibitem{Rubakov:2014jja}
V.~A. Rubakov, ``{The Null Energy Condition and its violation},'' {\em Phys.
  Usp.}, vol.~57, pp.~128--142, 2014.
\newblock [Usp. Fiz. Nauk184,no.2,137(2014)].

\bibitem{Dolan:2016bxj}
S.~R. Dolan and J.~O. Shipley, ``{Stable photon orbits in stationary
  axisymmetric electrovacuum spacetimes},'' {\em Phys. Rev.}, vol.~D94, no.~4,
  p.~044038, 2016.

\bibitem{Kerr:1963ud}
R.~P. Kerr, ``{Gravitational field of a spinning mass as an example of
  algebraically special metrics},'' {\em Phys.Rev.Lett.}, vol.~11,
  pp.~237--238, 1963.

\bibitem{Robinsoon:2004zz}
D.~Robinson, ``Four decades of black holes uniqueness theorems,'' in {\em The
  Kerr Spacetime: Rotating Black Holes in General Relativity} (D.~Wiltshire,
  M.~Visser, and S.~M. Scott, eds.), Cambridge University Press, 2009.

\bibitem{Chrusciel:2012jk}
P.~T. Chrusciel, J.~L. Costa, and M.~Heusler, ``{Stationary Black Holes:
  Uniqueness and Beyond},'' {\em Living Rev.Rel.}, vol.~15, p.~7, 2012.

\bibitem{Heusler:1998ua}
M.~Heusler, ``{Stationary black holes: Uniqueness and beyond},'' {\em Living
  Rev.Rel.}, vol.~1, p.~6, 1998.

\bibitem{Wilkins:1972rs}
D.~C. Wilkins, ``{Bound Geodesics in the Kerr Metric},'' {\em Phys. Rev.},
  vol.~D5, pp.~814--822, 1972.

\bibitem{nickalls_2006}
R.~W.~D. Nickalls, ``Viète, descartes and the cubic equation,'' {\em The
  Mathematical Gazette}, vol.~90, no.~518, p.~203–208, 2006.

\bibitem{nickalls1993new}
R.~W. Nickalls, ``A new approach to solving the cubic: Cardan’s solution
  revealed,'' {\em The Mathematical Gazette}, vol.~77, no.~480, pp.~354--359,
  1993.

\bibitem{Zakharov:2005ek}
A.~F. Zakharov, F.~De~Paolis, G.~Ingrosso, and A.~A. Nucita, ``{Measuring the
  black hole parameters in the galactic center with RADIOASTRON},'' {\em New
  Astron.}, vol.~10, pp.~479--489, 2005.

\bibitem{Herdeiro:2015gia}
C.~Herdeiro and E.~Radu, ``{Construction and physical properties of Kerr black
  holes with scalar hair},'' {\em Class.Quant.Grav.}, vol.~32, no.~14,
  p.~144001, 2015.

\bibitem{Herdeiro:2014goa}
C.~A.~R. Herdeiro and E.~Radu, ``{Kerr black holes with scalar hair},'' {\em
  Phys.Rev.Lett.}, vol.~112, p.~221101, 2014.

\bibitem{Muhleman:1970zz}
D.~O. Muhleman, R.~D. Ekers, and E.~B. Fomalont, ``{Radio Interferometric Test
  of the General Relativistic Light Bending Near the Sun},'' {\em Phys. Rev.
  Lett.}, vol.~24, pp.~1377--1380, 1970.

\bibitem{perlick2000ray}
V.~Perlick, {\em Ray Optics, Fermat’s Principle, and Applications to General
  Relativity}.
\newblock Lecture Notes in Physics Monographs, Springer, 2000.

\bibitem{Tsupko:2013cqa}
O.~{\relax Yu}. Tsupko and G.~S. Bisnovatyi-Kogan, ``{Gravitational lensing in
  plasma: Relativistic images at homogeneous plasma},'' {\em Phys. Rev.},
  vol.~D87, no.~12, p.~124009, 2013.

\bibitem{Bisnovatyi-Kogan:2015dxa}
G.~S. Bisnovatyi-Kogan and O.~{\relax Yu}. Tsupko, ``{Gravitational Lensing in
  Plasmic Medium},'' 2015.
\newblock [Plasma Phys. Rep.41,562(2015)].

\bibitem{Abdujabbarov:2016efm}
A.~Abdujabbarov, B.~Juraev, B.~Ahmedov, and Z.~Stuchlík, ``{Shadow of rotating
  wormhole in plasma environment},'' {\em Astrophys. Space Sci.}, vol.~361,
  no.~7, p.~226, 2016.

\bibitem{Abdujabbarov:2015pqp}
A.~Abdujabbarov, B.~Toshmatov, Z.~Stuchlík, and B.~Ahmedov, ``{Shadow of the
  rotating black hole with quintessential energy in the presence of plasma},''
  {\em Int. J. Mod. Phys.}, vol.~D26, no.~06, p.~1750051, 2016.

\bibitem{Perlick:2017fio}
V.~Perlick and O.~{\relax Yu}. Tsupko, ``{Light propagation in a plasma on Kerr
  spacetime: Separation of the Hamilton-Jacobi equation and calculation of the
  shadow},'' {\em Phys. Rev.}, vol.~D95, no.~10, p.~104003, 2017.

\bibitem{Herdeiro:2016tmi}
C.~Herdeiro, E.~Radu, and H.~Runarsson, ``{Kerr black holes with Proca hair},''
  {\em Class. Quant. Grav.}, vol.~33, no.~15, p.~154001, 2016.

\bibitem{East:2017ovw}
W.~E. East and F.~Pretorius, ``{Superradiant Instability and Backreaction of
  Massive Vector Fields around Kerr Black Holes},'' {\em Phys. Rev. Lett.},
  vol.~119, no.~4, p.~041101, 2017.

\bibitem{Herdeiro:2017phl}
C.~A.~R. Herdeiro and E.~Radu, ``{Kerr black holes with synchronised hair: an
  analytic model and dynamical formation},'' 2017.

\bibitem{Hod:2012px}
S.~Hod, ``{Stationary Scalar Clouds Around Rotating Black Holes},'' {\em
  Phys.Rev.}, vol.~D86, p.~104026, 2012.

\bibitem{Hod:2013zza}
S.~Hod, ``{Stationary resonances of rapidly-rotating Kerr black holes},'' {\em
  The European Physical Journal C 73,}, vol.~2378, 2013.

\bibitem{Perlick:2004tq}
V.~Perlick, ``{Gravitational lensing from a spacetime perspective},'' {\em
  Living Rev. Rel.}, vol.~7, p.~9, 2004.

\bibitem{Bohn:2014xxa}
A.~Bohn, W.~Throwe, F.~H{\'e}bert, K.~Henriksson, D.~Bunandar, {\em et~al.},
  ``{What does a binary black hole merger look like?},'' {\em
  Class.Quant.Grav.}, vol.~32, no.~6, p.~065002, 2015.

\bibitem{Cunha:2015yba}
P.~V.~P. Cunha, C.~A.~R. Herdeiro, E.~Radu, and H.~F. Runarsson, ``{Shadows of
  Kerr black holes with scalar hair},'' {\em Phys. Rev. Lett.}, vol.~115,
  no.~21, p.~211102, 2015.

\bibitem{Wang:2017hjl}
M.~Wang, S.~Chen, and J.~Jing, ``{Shadow casted by a Konoplya-Zhidenko rotating
  non-Kerr black hole},'' {\em JCAP}, vol.~1710, no.~10, p.~051, 2017.

\bibitem{jose1998classical}
J.~Jos{\'e} and E.~Saletan, {\em Classical Dynamics: A Contemporary Approach}.
\newblock Cambridge University Press, 1998.

\bibitem{Brito:2015pxa}
R.~Brito, V.~Cardoso, C.~A.~R. Herdeiro, and E.~Radu, ``{Proca stars:
  Gravitating Bose-Einstein condensates of massive spin 1 particles},'' {\em
  Phys. Lett.}, vol.~B752, pp.~291--295, 2016.

\bibitem{1974SvA....17..562P}
S.~B. {Pikel'Ner}, ``{Book Review: Ya. B. Zel'dovich and I. D. Novikov. The
  theory of gravitation and stellar evolution},'' {\em Soviet Astronomy},
  vol.~17, p.~562, Feb. 1974.

\bibitem{Chakraborty:2016lxo}
S.~Chakraborty and S.~SenGupta, ``{Strong gravitational lensing --- A probe for
  extra dimensions and Kalb-Ramond field},'' {\em JCAP}, vol.~1707, no.~07,
  p.~045, 2017.

\bibitem{Cunha:2016wzk}
P.~V.~P. Cunha, C.~A.~R. Herdeiro, B.~Kleihaus, J.~Kunz, and E.~Radu,
  ``{Shadows of Einstein-dilaton-Gauss-Bonnet black holes},'' {\em Phys.
  Lett.}, vol.~B768, pp.~373--379, 2017.

\bibitem{Ostrogradsky:1850fid}
M.~Ostrogradsky, ``{M{\'e}moires sur les {\'e}quations diff{\'e}rentielles,
  relatives au probl{\`e}me des isop{\'e}rim{\`e}tres},'' {\em Mem. Acad. St.
  Petersbourg}, vol.~6, no.~4, pp.~385--517, 1850.

\bibitem{Lovelock:1971yv}
D.~Lovelock, ``{The Einstein tensor and its generalizations},'' {\em J. Math.
  Phys.}, vol.~12, pp.~498--501, 1971.

\bibitem{Zwiebach:1985uq}
B.~Zwiebach, ``{Curvature Squared Terms and String Theories},'' {\em Phys.
  Lett.}, vol.~156B, pp.~315--317, 1985.

\bibitem{Kanti:1995vq}
P.~Kanti, N.~E. Mavromatos, J.~Rizos, K.~Tamvakis, and E.~Winstanley,
  ``{Dilatonic black holes in higher curvature string gravity},'' {\em Phys.
  Rev.}, vol.~D54, pp.~5049--5058, 1996.

\bibitem{Kanti:1997br}
P.~Kanti, N.~E. Mavromatos, J.~Rizos, K.~Tamvakis, and E.~Winstanley,
  ``{Dilatonic black holes in higher curvature string gravity. 2: Linear
  stability},'' {\em Phys. Rev.}, vol.~D57, pp.~6255--6264, 1998.

\bibitem{Torii:1996yi}
T.~Torii, H.~Yajima, and K.-i. Maeda, ``{Dilatonic black holes with
  Gauss-Bonnet term},'' {\em Phys. Rev.}, vol.~D55, pp.~739--753, 1997.

\bibitem{Alexeev:1996vs}
S.~O. Alexeev and M.~V. Pomazanov, ``{Black hole solutions with dilatonic hair
  in higher curvature gravity},'' {\em Phys. Rev.}, vol.~D55, pp.~2110--2118,
  1997.

\bibitem{Melis:2005ji}
M.~Melis and S.~Mignemi, ``{Global properties of charged dilatonic Gauss-Bonnet
  black holes},'' {\em Phys. Rev.}, vol.~D73, p.~083010, 2006.

\bibitem{Chen:2006ge}
C.-M. Chen, D.~V. Gal'tsov, and D.~G. Orlov, ``{Extremal black holes in D=4
  Gauss-Bonnet gravity},'' {\em Phys. Rev.}, vol.~D75, p.~084030, 2007.

\bibitem{Chen:2008hk}
C.-M. Chen, D.~V. Gal'tsov, and D.~G. Orlov, ``{Extremal dyonic black holes in
  D=4 Gauss-Bonnet gravity},'' {\em Phys. Rev.}, vol.~D78, p.~104013, 2008.

\bibitem{Kleihaus:2011tg}
B.~Kleihaus, J.~Kunz, and E.~Radu, ``{Rotating Black Holes in Dilatonic
  Einstein-Gauss-Bonnet Theory},'' {\em Phys. Rev. Lett.}, vol.~106, p.~151104,
  2011.

\bibitem{Kleihaus:2015aje}
B.~Kleihaus, J.~Kunz, S.~Mojica, and E.~Radu, ``{Spinning black holes in
  Einstein–Gauss-Bonnet–dilaton theory: Nonperturbative solutions},'' {\em
  Phys. Rev.}, vol.~D93, no.~4, p.~044047, 2016.

\bibitem{Pani:2009wy}
P.~Pani and V.~Cardoso, ``{Are black holes in alternative theories serious
  astrophysical candidates? The Case for Einstein-Dilaton-Gauss-Bonnet black
  holes},'' {\em Phys. Rev.}, vol.~D79, p.~084031, 2009.

\bibitem{Pani:2011gy}
P.~Pani, C.~F.~B. Macedo, L.~C.~B. Crispino, and V.~Cardoso, ``{Slowly rotating
  black holes in alternative theories of gravity},'' {\em Phys. Rev.},
  vol.~D84, p.~087501, 2011.

\bibitem{Ayzenberg:2014aka}
D.~Ayzenberg and N.~Yunes, ``{Slowly-Rotating Black Holes in
  Einstein-Dilaton-Gauss-Bonnet Gravity: Quadratic Order in Spin Solutions},''
  {\em Phys. Rev.}, vol.~D90, p.~044066, 2014.
\newblock [Erratum: Phys. Rev.D91,no.6,069905(2015)].

\bibitem{Maselli:2015tta}
A.~Maselli, P.~Pani, L.~Gualtieri, and V.~Ferrari, ``{Rotating black holes in
  Einstein-Dilaton-Gauss-Bonnet gravity with finite coupling},'' {\em Phys.
  Rev.}, vol.~D92, no.~8, p.~083014, 2015.

\bibitem{Herdeiro:2015waa}
C.~A.~R. Herdeiro and E.~Radu, ``{Asymptotically flat black holes with scalar
  hair: a review},'' {\em Int. J. Mod. Phys.}, vol.~D24, no.~09, p.~1542014,
  2015.

\bibitem{Amarilla:2011fx}
L.~Amarilla and E.~F. Eiroa, ``{Shadow of a rotating braneworld black hole},''
  {\em Phys. Rev.}, vol.~D85, p.~064019, 2012.

\bibitem{Yumoto:2012kz}
A.~Yumoto, D.~Nitta, T.~Chiba, and N.~Sugiyama, ``{Shadows of Multi-Black
  Holes: Analytic Exploration},'' {\em Phys. Rev.}, vol.~D86, p.~103001, 2012.

\bibitem{Abdujabbarov:2012bn}
A.~Abdujabbarov, F.~Atamurotov, Y.~Kucukakca, B.~Ahmedov, and U.~Camci,
  ``{Shadow of Kerr-Taub-NUT black hole},'' {\em Astrophys. Space Sci.},
  vol.~344, pp.~429--435, 2013.

\bibitem{Amarilla:2013sj}
L.~Amarilla and E.~F. Eiroa, ``{Shadow of a Kaluza-Klein rotating dilaton black
  hole},'' {\em Phys. Rev.}, vol.~D87, no.~4, p.~044057, 2013.

\bibitem{Nedkova:2013msa}
P.~G. Nedkova, V.~K. Tinchev, and S.~S. Yazadjiev, ``{Shadow of a rotating
  traversable wormhole},'' {\em Phys. Rev.}, vol.~D88, no.~12, p.~124019, 2013.

\bibitem{Atamurotov:2013dpa}
F.~Atamurotov, A.~Abdujabbarov, and B.~Ahmedov, ``{Shadow of rotating
  Hořava-Lifshitz black hole},'' {\em Astrophys. Space Sci.}, vol.~348,
  pp.~179--188, 2013.

\bibitem{Atamurotov:2013sca}
F.~Atamurotov, A.~Abdujabbarov, and B.~Ahmedov, ``{Shadow of rotating non-Kerr
  black hole},'' {\em Phys. Rev.}, vol.~D88, no.~6, p.~064004, 2013.

\bibitem{Li:2013jra}
Z.~Li and C.~Bambi, ``{Measuring the Kerr spin parameter of regular black holes
  from their shadow},'' {\em JCAP}, vol.~1401, p.~041, 2014.

\bibitem{Tinchev:2013nba}
V.~K. Tinchev and S.~S. Yazadjiev, ``{Possible imprints of cosmic strings in
  the shadows of galactic black holes},'' {\em Int. J. Mod. Phys.}, vol.~D23,
  p.~1450060, 2014.

\bibitem{Wei:2013kza}
S.-W. Wei and Y.-X. Liu, ``{Observing the shadow of
  Einstein-Maxwell-Dilaton-Axion black hole},'' {\em JCAP}, vol.~1311, p.~063,
  2013.

\bibitem{Tsukamoto:2014tja}
N.~Tsukamoto, Z.~Li, and C.~Bambi, ``{Constraining the spin and the deformation
  parameters from the black hole shadow},'' {\em JCAP}, vol.~1406, p.~043,
  2014.

\bibitem{Grenzebach:2014fha}
A.~Grenzebach, V.~Perlick, and C.~Lammerzahl, ``{Photon Regions and Shadows of
  Kerr-Newman-NUT Black Holes with a Cosmological Constant},'' {\em Phys.
  Rev.}, vol.~D89, no.~12, p.~124004, 2014.

\bibitem{Lu:2014zja}
R.-S. Lu, A.~E. Broderick, F.~Baron, J.~D. Monnier, V.~L. Fish, S.~S. Doeleman,
  and V.~Pankratius, ``{Imaging the Supermassive Black Hole Shadow and Jet Base
  of M87 with the Event Horizon Telescope},'' {\em Astrophys. J.}, vol.~788,
  p.~120, 2014.

\bibitem{Papnoi:2014aaa}
U.~Papnoi, F.~Atamurotov, S.~G. Ghosh, and B.~Ahmedov, ``{Shadow of
  five-dimensional rotating Myers-Perry black hole},'' {\em Phys. Rev.},
  vol.~D90, no.~2, p.~024073, 2014.

\bibitem{Sakai:2014pga}
N.~Sakai, H.~Saida, and T.~Tamaki, ``{Gravastar Shadows},'' {\em Phys. Rev.},
  vol.~D90, no.~10, p.~104013, 2014.

\bibitem{Psaltis:2014mca}
D.~Psaltis, F.~Ozel, C.-K. Chan, and D.~P. Marrone, ``{A General Relativistic
  Null Hypothesis Test with Event Horizon Telescope Observations of the
  black-hole shadow in Sgr A*},'' {\em Astrophys. J.}, vol.~814, no.~2, p.~115,
  2015.

\bibitem{Wei:2015dua}
S.-W. Wei, P.~Cheng, Y.~Zhong, and X.-N. Zhou, ``{Shadow of noncommutative
  geometry inspired black hole},'' {\em JCAP}, vol.~1508, no.~08, p.~004, 2015.

\bibitem{Abdolrahimi:2015rua}
S.~Abdolrahimi, R.~B. Mann, and C.~Tzounis, ``{Distorted Local Shadows},'' {\em
  Phys. Rev.}, vol.~D91, no.~8, p.~084052, 2015.

\bibitem{Moffat:2015kva}
J.~W. Moffat, ``{Modified Gravity Black Holes and their Observable Shadows},''
  {\em Eur. Phys. J.}, vol.~C75, no.~3, p.~130, 2015.

\bibitem{Grenzebach:2015uva}
A.~Grenzebach, ``{Aberrational Effects for Shadows of Black Holes},'' {\em
  Fund. Theor. Phys.}, vol.~179, pp.~823--832, 2015.

\bibitem{Vincent:2015xta}
F.~H. Vincent, Z.~Meliani, P.~Grandclement, E.~Gourgoulhon, and O.~Straub,
  ``{Imaging a boson star at the Galactic center},'' {\em Class. Quant. Grav.},
  vol.~33, no.~10, p.~105015, 2016.

\bibitem{Grenzebach:2015oea}
A.~Grenzebach, V.~Perlick, and C.~Lammerzahl, ``{Photon Regions and Shadows of
  Accelerated Black Holes},'' 2015.

\bibitem{Abdujabbarov:2015xqa}
A.~A. Abdujabbarov, L.~Rezzolla, and B.~J. Ahmedov, ``{A coordinate-independent
  characterization of a black hole shadow},'' {\em Mon. Not. Roy. Astron.
  Soc.}, vol.~454, no.~3, pp.~2423--2435, 2015.

\bibitem{Ortiz:2015rma}
N.~Ortiz, O.~Sarbach, and T.~Zannias, ``{Shadow of a naked singularity},'' {\em
  Phys. Rev.}, vol.~D92, no.~4, p.~044035, 2015.

\bibitem{Ghasemi-Nodehi:2015raa}
M.~Ghasemi-Nodehi, Z.~Li, and C.~Bambi, ``{Shadows of CPR black holes and tests
  of the Kerr metric},'' {\em Eur. Phys. J.}, vol.~C75, p.~315, 2015.

\bibitem{Ohgami:2015nra}
T.~Ohgami and N.~Sakai, ``{Wormhole shadows},'' {\em Phys. Rev.}, vol.~D91,
  no.~12, p.~124020, 2015.

\bibitem{Atamurotov:2015xfa}
F.~Atamurotov, S.~G. Ghosh, and B.~Ahmedov, ``{Horizon structure of rotating
  Einstein–Born–Infeld black holes and shadow},'' {\em Eur. Phys. J.},
  vol.~C76, no.~5, p.~273, 2016.

\bibitem{Perlick:2015vta}
V.~Perlick, O.~{\relax Yu}. Tsupko, and G.~S. Bisnovatyi-Kogan, ``{Influence of
  a plasma on the shadow of a spherically symmetric black hole},'' {\em Phys.
  Rev.}, vol.~D92, no.~10, p.~104031, 2015.

\bibitem{Bambi:2015rda}
C.~Bambi, ``{Testing the Kerr Paradigm with the Black Hole Shadow},'' in {\em
  {14th Marcel Grossmann Meeting on General Relativity on Recent Developments
  in Theoretical and Experimental General Relativity, Astrophysics, and
  Relativistic Field Theories (MG14) Rome, Italy, July 12-18, 2015}}, 2015.

\bibitem{Atamurotov:2015nra}
F.~Atamurotov and B.~Ahmedov, ``{Optical properties of black hole in the
  presence of plasma: shadow},'' {\em Phys. Rev.}, vol.~D92, p.~084005, 2015.

\bibitem{Yang:2015hwf}
L.~Yang and Z.~Li, ``{Shadow of a dressed black hole and determination of spin
  and viewing angle},'' {\em Int. J. Mod. Phys.}, vol.~D25, no.~02, p.~1650026,
  2015.

\bibitem{Tinchev:2015apf}
V.~K. Tinchev, ``{The Shadow of Generalized Kerr Black Holes with Exotic
  Matter},'' {\em Chin. J. Phys.}, vol.~53, p.~110113, 2015.

\bibitem{Amir:2016cen}
M.~Amir and S.~G. Ghosh, ``{Shapes of rotating nonsingular black hole
  shadows},'' {\em Phys. Rev.}, vol.~D94, no.~2, p.~024054, 2016.

\bibitem{Johannsen:2015hib}
T.~Johannsen, A.~E. Broderick, P.~M. Plewa, S.~Chatzopoulos, S.~S. Doeleman,
  F.~Eisenhauer, V.~L. Fish, R.~Genzel, O.~Gerhard, and M.~D. Johnson,
  ``{Testing General Relativity with the Shadow Size of Sgr A*},'' {\em Phys.
  Rev. Lett.}, vol.~116, no.~3, p.~031101, 2016.

\bibitem{Abdujabbarov:2016hnw}
A.~Abdujabbarov, M.~Amir, B.~Ahmedov, and S.~G. Ghosh, ``{Shadow of rotating
  regular black holes},'' {\em Phys. Rev.}, vol.~D93, no.~10, p.~104004, 2016.

\bibitem{Cunha:2016bpi}
P.~V.~P. Cunha, C.~A.~R. Herdeiro, E.~Radu, and H.~F. Runarsson, ``{Shadows of
  Kerr black holes with and without scalar hair},'' in {\em {3rd Amazonian
  Symposium on Physics and 5th NRHEP Network Meeting is approaching:
  Celebrating 100 Years of General Relativity Belem, Brazil, September
  28-October 2, 2015}}, 2016.

\bibitem{Huang:2016qnl}
Y.~Huang, S.~Chen, and J.~Jing, ``{Double shadow of a regular phantom black
  hole as photons couple to Weyl tensor},'' 2016.

\bibitem{Dastan:2016vhb}
S.~Dastan, R.~Saffari, and S.~Soroushfar, ``{Shadow of a Charged Rotating Black
  Hole in $f(R)$ Gravity},'' 2016.

\bibitem{Younsi:2016azx}
Z.~Younsi, A.~Zhidenko, L.~Rezzolla, R.~Konoplya, and Y.~Mizuno, ``{A new
  method for shadow calculations: application to parameterised axisymmetric
  black holes},'' 2016.

\bibitem{Ohgami:2016iqm}
T.~Ohgami and N.~Sakai, ``{Wormhole shadows in rotating dust},'' {\em Phys.
  Rev.}, vol.~D94, no.~6, p.~064071, 2016.

\bibitem{Mureika:2016efo}
J.~R. Mureika and G.~U. Varieschi, ``{Black hole shadows in fourth-order
  conformal Weyl gravity},'' 2016.

\bibitem{Sharif:2016znp}
M.~Sharif and S.~Iftikhar, ``{Shadow of a Charged Rotating Non-Commutative
  Black Hole},'' {\em Eur. Phys. J.}, vol.~C76, no.~11, p.~630, 2016.

\bibitem{Tsupko:2017rdo}
O.~{\relax Yu}. Tsupko, ``{Analytical calculation of black hole spin using
  deformation of the shadow},'' {\em Phys. Rev.}, vol.~D95, no.~10, p.~104058,
  2017.

\bibitem{Bisnovatyi-Kogan:2017kii}
G.~Bisnovatyi-Kogan and O.~Tsupko, ``{Gravitational Lensing in Presence of
  Plasma: Strong Lens Systems, Black Hole Lensing and Shadow},'' {\em
  Universe}, vol.~3, no.~3, p.~57, 2017.

\bibitem{Amir:2017slq}
M.~Amir, B.~P. Singh, and S.~G. Ghosh, ``{Shadows of rotating five-dimensional
  EMCS black holes},'' 2017.

\bibitem{Alhamzawi:2017iyn}
A.~Alhamzawi, ``{Observing the shadow of modified gravity black hole},'' {\em
  Int. J. Mod. Phys.}, vol.~D26, no.~14, p.~1750156, 2017.

\bibitem{Tsukamoto:2017fxq}
N.~Tsukamoto, ``{Black hole shadow in an asymptotically-flat, stationary, and
  axisymmetric spacetime: the Kerr-Newman and rotating regular black holes},''
  2017.

\bibitem{Mars:2017jkk}
M.~Mars, C.~F. Paganini, and M.~A. Oancea, ``{The fingerprints of black
  holes-shadows and their degeneracies},'' {\em Class. Quant. Grav.}, vol.~35,
  no.~2, p.~025005, 2018.

\bibitem{Wang:2017qhh}
M.~Wang, S.~Chen, and J.~Jing, ``{Shadows of a compact object with magnetic
  dipole by chaotic lensing},'' 2017.

\bibitem{Singh:2017xle}
B.~P. Singh, ``{Rotating charge black holes shadow in quintessence},'' 2017.

\bibitem{Eiroa:2017uuq}
E.~F. Eiroa and C.~M. Sendra, ``{Shadow cast by rotating braneworld black holes
  with a cosmological constant},'' {\em Eur. Phys. J.}, vol.~C78, no.~2, p.~91,
  2018.

\end{thebibliography}
\bibliographystyle{ieeetr}

\end{document}